\documentclass{iopart}
\usepackage{graphicx}  
\usepackage{latexsym}
\usepackage{amssymb}

\jl{6}        
\eqnobysec    

\def\beq{\begin{equation}}
\def\eeq{\end{equation}}

\def\rmd{{\rm d}}
\def\rmD{{\rm D}}

\def\rightcontract{\mathop{\hbox{\vrule width0.5pt height6pt%
  \vrule height0.5pt width6pt}}}

\begin{document}

\title[Strains in General Relativity]
{Strains in General Relativity}

\author{
Donato Bini$^* {}^\S{}^\P$,
Fernando de Felice$^\dagger$ 
and 
Andrea Geralico$^\S$
}

\address{
  ${}^*$\
  Istituto per le Applicazioni del Calcolo ``M. Picone,'' CNR I-00161 Rome, Italy
}
\address{
  ${}^\S$\
  International Center for Relativistic Astrophysics - ICRA,
  University of Rome ``La Sapienza,'' I-00185 Rome, Italy
}
\address{
  ${}^\P$
  INFN - Sezione di Firenze, Polo Scientifico, Via Sansone 1, 
  I-50019, Sesto Fiorentino (FI), Italy 
}
\address{
  ${}^\dagger$\
  Dipartimento di Fisica, Universit\`a di Padova, and INFN, Sezione di Padova, Via Marzolo 8,  I-35131 Padova, Italy
}


\begin{abstract}
The definition of relative accelerations and strains among a set of 
comoving particles is studied in connection with  the geometric 
properties of the  frame adapted to a \lq\lq fiducial observer." We find 
that a relativistically complete and correct definition of strains must 
take into account 
the transport law of the chosen spatial triad along the observer's 
congruence. We use special congruences of (accelerated) test particles 
in some familiar spacetimes to elucidate such a point. 
The celebrated idea of Szekeres' compass of inertia,  arising when 
studying geodesic deviation among a set of free-falling particles, is 
here generalized 
to the case of accelerated particles.
In doing so we have naturally contributed to the theory of relativistic gravity gradiometer.
Moreover, our analysis was made in an observer-dependent form, a fact that
would be very useful when thinking about general 
relativistic tests on space stations orbiting compact objects like black holes and 
also in other interesting gravitational situations.
\end{abstract}

\pacno{04.20.Cv}

\section{Introduction}

In a series of papers, long ago, de Felice and coworkers \cite{fdf1,fdf2,fdf3,sem-fdf} defined and studied the relative strains among a set of comoving particles in black hole spacetimes, confined to a normal neighborhood of the observer's world line and in a well specified state of motion. 
The particles were considered as test with respect to the background geometry.

Starting from that analysis, we consider here how the definition of relative accelerations and strains is affected by the geometric properties of the  frame adapted to the fiducial observer (e.g. transport law of the spatial triad along the observer's congruence). 
Within this more general context we reconsider previous works and extend that discussion in view of possible quasi-local experiments in space laboratories moving in different gravitational environments.

The basis of our observer-dependent analysis of relative strains is the concept of gravitational compass introduced by Szekeres \cite{szekeres} 
(see also \cite{pir1,pir2,ciu1,ciu2,aud-lae}) and the related discussion about the problem of setting up a preferred frame within which to study the gravitational field. 
According to Szekeres, a gravitational compass consists in an arrangement of three
test particles joined by springs to a central observer; their relative deviation is then investigated via the geodesic deviation equation
to deduce the physical significance of the Weyl tensor components. At the instant of measurement the reference particle drops the apparatus observing the strains on the springs. The relative acceleration between two nearby geodesics is completely determined by the electric part of the Riemann tensor, which can be thought of as a symmetric force distribution whose six independent components are the strains on the six springs. When the off diagonal terms (i.e. the transverse strains) vanish, the springs connecting the test particles to the observer lie along the principal axes of the tidal force matrix, so that the apparatus maps out the local gravitational field, acting just as a sort of compass.
Expressing the Riemann tensor in terms of the Weyl tensor (with which coincides in the case of vacuum spacetimes) allowed Szekeres to discuss the physical meaning of the Petrov classification. The electric part of the Weyl tensor represents the only direct curvature contribution to the geodesic deviation equation (as well as the deviation equation for general non-geodesic motion), introducing shearing forces due to its property of being symmetric and trace-free. 
Actually Szekeres' gravitational compass is only valid to describe an idealized situation. 
For any practical use, in fact, it should be replaced by a \lq\lq gravity gradiometer," i.e. a device to perform measurements of the local gradient of  the tidal gravitational force. The theory of a relativistic gravity gradiometer has been developed by Mashhoon, Theiss, Paik and  Will \cite{mas-the,mas-pai-wil} in view of satellite experiments around the Earth in the framework of Post-Newtonian approximation. It should also be noted that a modern observational trend is to use atomic interferometry to build the future  
generation of highly precise gravity gradiometers (see \cite{matsko} and references therein).

Recently Chicone and Mashhoon \cite{chicmash1} obtained a generalized geodesic deviation equation in Fermi coordinates as well 
as in arbitrary coordinates as a Taylor expansion in powers of the components of the 
deviation vector, retaining terms up to first order, but without any restriction on the 
relative spatial velocities. They then investigated in a number of papers \cite{chicmash2,chicmash3,chicmash4} the motion of a swarm of free particles (in both non-relativistic 
and relativistic regime) relative to a free reference particle which is on a radial escape 
trajectory away from a collapsed object (a Schwarzschild as well as a Kerr black hole), discussing the astrophysical implications of the related (observer-dependent) tidal acceleration mechanism. 
The further dependence of the deviation equation on the four acceleration of the observer as well as his three velocity
has been accounted very recently by Mullari and Tammelo \cite{mullari}.

Aim of this work is to set up the necessary assessment to assure stability of extended bodies as they move in a given spacetime, identifying which parts of them should be provided of a more rigid structure to resist tidal or acceleration strains, according to their internal constituency. In fact we are not concerned here with internal stresses but with the external field of strains generated by the geometrical environment.
Special attention will be devoted to type D vacuum stationary axisymmetric spacetimes. 
In this case, Szekeres' analysis suggests that the tidal strains cause the distorsion of a sphere of test free particles about the observer into an ellipsoid, as a typical behaviour of particles falling towards the central attracting body. 
We are thus interested in studying how Szekeres' picture as well as the one associated with the relativistic gravity gradiometry modifies when the acceleration strains are also present, and which frame is most convenient to measure either tidal or inertial forces experienced by an extended body.

The paper is organized as follows. In Section 2 we derive the relative acceleration equation and give the (observer and frame-dependent) definition of the strain tensor. In Section 3 we analyze the strains which affect a bunch of uniformly rotating particles in the flat Minkowski spacetime. 
In Section \ref{kerrcase} we discuss instead the deviation acceleration and associated strains in vacuum stationary axisymmetric spacetimes from either geometrically or physically motivated timelike congruences. In particular we explore the cases of the Born-rigid congruence of static observers, of the irrotational family of Zero Angular Momentum Observers (ZAMOs), and (limiting to the Schwarzschild and Kerr black hole spacetimes) of the geodesic and irrotational congruence of Painlev\'e-Gullstrand \cite{doran} orbits. 

In what follows greek indices refer to coordinate components while latin indices  refer to tetrad components. The formers run from 0 to 3, the latters from 1 to 3. The index (apex) 0  is often used with the meaning of time tetrad component too. Units are choosen so that $c=G=1$ and the adopted metric signature is $-+++$.

\section{The relative acceleration equation}

Let us consider a bunch of test particles, i.e. a congruence ${\mathcal C}_U$ of timelike world lines, with unit tangent vector
$U$ ($U\cdot U=-1$)  parametrized by the proper time $\tau_U$. 
Let ${\mathcal C}_*$ be the reference world line of the congruence,  which we consider as that of the \lq\lq fiducial observer." 
In general, the lines of the congruence ${\mathcal C}$ as well as that of the observer are accelerated with  acceleration   
$a(U)=\nabla_U U$. 

The separation  between the line ${\mathcal C}_*$ and a general line of the congruence is represented by a connecting vector $Y$, i.e. a vector undergoing Lie transport along $U$:
\beq
\label{eq:1}
\pounds_U Y=0\  \qquad \rightarrow \qquad \nabla_UY=\nabla_YU\ .
\eeq
The term $\nabla_YU$ in Eq.~(\ref{eq:1})$_2$ involves the covariant derivative of $U$, which can be written in terms of the kinematical fields of the congruence as follows:
\beq
\label{eq:1b}
\nabla_\alpha U^\beta = - a(U)^\beta U_\alpha -K(U)^\beta{}_\alpha\ ,
\eeq
where $K(U)^\beta{}_\alpha=\omega(U)^\beta{}_\alpha -\theta(U)^\beta{}_\alpha$ is the kinematical tensor, which summarizes the vorticity of the congruence $\omega(U)_{\alpha\beta}=K(U)_{[\alpha\beta]}$ and the expansion $\theta(U)_{\alpha\beta}=-K(U)_{(\alpha\beta)}$. Here square and round brakets denote antisymmetrization and symmetrization of tensor indices, respectively. 
From Eq. (\ref{eq:1b}) we also have
\begin{eqnarray}
\fl\quad
\omega(U)_{\alpha\beta} = P(U)_\alpha^\mu P(U)_\beta^\nu \nabla_{[\mu} U_{\nu]}\ , \qquad
\theta(U)_{\alpha\beta} = P(U)_\alpha^\mu P(U)_\beta^\nu \nabla_{(\mu}U_{\nu)}\ ,
\end{eqnarray}
where $P(U)_\alpha^\beta=\delta_\alpha^\beta+U_\alpha U^\beta$ is the operator of the orthogonal projection  with respect  to $U$. 
Thus Eq. (\ref{eq:1}) becomes
\beq
\frac{\rmD Y}{\rmd \tau_U }=\nabla_Y U= -(Y\cdot U)a(U)-K(U)\rightcontract Y\ ,
\eeq
where $\rightcontract$ denotes index contraction \cite{mfg}, so that
$[K(U)\rightcontract Y]^\beta=K(U)^\beta{}_\alpha Y^\alpha$.

The covariant derivative along $U$ of both sides of Eq.~(\ref{eq:1})$_2$ gives rise to the \lq\lq relative acceleration equation"
\beq
\label{eq:2}
\frac{\rmD^2 Y}{\rmd \tau_U^2 }=-R(U,Y)U+\nabla_Y a(U)\ ,
\eeq
where $R(U,Y)U\equiv R^\alpha{}_{\beta\gamma\delta}U^\beta Y^\gamma U^\delta$ represents the tidal force contribution to the relative acceleration, whereas $\nabla_Y a(U)$ is the \lq\lq inertial"  contribution due to the observer's acceleration.
We notice that $R^\alpha{}_{\beta\gamma\delta}U^\beta U^\delta={\mathcal E}(U)^\alpha{}_\gamma$ is the electric part of the Riemann tensor (as measured by the observer $U$).
Equation (\ref{eq:2}) can then be conveniently rewritten as follows:
\beq
\label{eq:3}
\frac{\rmD^2 Y}{\rmd \tau_U^2 }=-\Omega \rightcontract Y\ , \qquad \Omega ={\mathcal E}(U)-\nabla a(U)\ .
\eeq

Let us now set up an orthonormal frame $\{E_\alpha\}=\{U\equiv E_0,E_a\}$ adapted to the congruence $U$ and write both the Lie transport equation (\ref{eq:1})$_2$ and the relative acceleration equation (\ref{eq:2}) with respect to this frame. 
The spatial triad is generic in the sense that it rotates with a certain angular velocity $\omega_{({\rm fw},U,E)}$ with respect to gyro-fixed axes along $U$:
\beq
\label{spat_frame}
\fl\quad
P(U)\nabla_U E_a \equiv \nabla_{\rm (fw)}(U)E_a= \omega_{({\rm fw},U,E)}\times E_a\equiv C_{({\rm fw},U,E)}{}^b{}_a E_b\ .
\eeq
Here the subscript $({\rm fw},U,E)$ means that we are referring to a $U,E_a$ tetrad and a Fermi-Walker derivative operation;
the cross product $\times$ is defined in the local rest space of 
$U$ by $\eta(U)_{abc}=\eta_{\rho abc}U^\rho$ and the Fermi-Walker structure functions are defined by $C_{({\rm fw},U,E)}{}_{ab}=-\eta(U)_{abc}\,\omega_{({\rm fw},U,E)}^c$; $\nabla_{\rm (fw)}(U)$ is the {\it spatial Fermi-Walker derivative along $U$} \cite{mfg},  such that for a generic spatial vector $X$ (i.e. $X\cdot U=0$)
\beq
\label{spatfwder}
\nabla_{\rm (fw)}(U)X^a=\dot X^a + C_{ ({\rm fw},U,E)}{}^a{}_b\, X^b\ ,
\eeq
the overdot denoting differentiation with respect to the proper time $\tau_U$. 

Introduce the frame components of $Y$, i.e. the decomposition $Y=Y^0\,U+Y^a\,E_a$ (and the notation $\vec Y\equiv P(U)Y=Y^a\,E_a$, $\dot f= \rmd f/ \rmd \tau_U $). The Lie transport equation (\ref{eq:1})$_2$ then becomes
\begin{eqnarray}
\label{cond_lie}
\nabla_U Y&\equiv & \dot Y^0\, U+Y^0 a(U)+
\dot Y^a\, E_a +[\vec Y\cdot a(U)]U+\omega_{({\rm fw},U,E)}\times \vec Y  \nonumber \\
&=&
Y^0 a(U)-K(U)\rightcontract \vec Y\ ,
\end{eqnarray}
where the relation $Y_0=-Y^0$ has been used, whence
\beq
\left[\dot Y^0 +\vec Y\cdot a(U)\right]U+
\dot Y^a\, E_a +\omega_{({\rm fw},U,E)}\times \vec Y =-K(U)\rightcontract \vec Y\ ,
\eeq
yielding
\begin{eqnarray}
\label{conseq1}
&& \dot Y^0=-\vec Y\cdot a(U)\ , \\
\label{conseq2}
&& \dot Y^a +[\omega_{({\rm fw},U,E)}\times \vec Y]^a +K(U)^a{}_bY^b=0\ .
\end{eqnarray}
From the definition of $K(U)$
the \lq\lq relative velocity equation" (\ref{conseq2}) can be written as
\beq
\dot Y^a +[(\omega_{({\rm fw},U,E)}-\omega(U))\times \vec Y]^a -\theta(U)^a{}_bY^b=0\ ,
\eeq
implying that $\dot Y^a=0$ when $\omega_{({\rm fw},U,E)}=\omega(U)$ and $\theta(U)=0$. As we will see in Sec. 4.4 the latter condition is satisfied by 
a Frenet-Serret frame along a Born-rigid congruence.

Let us turn to the relative acceleration equation (\ref{eq:2}).
Substituting Eq.~(\ref{conseq1}) into the first line of (\ref{cond_lie}) leads to  
\beq
\nabla_U Y= Y^0 a(U)+ \dot Y^a\, E_a +\omega_{({\rm fw},U,E)}\times \vec Y\ .
\eeq
Taking the covariant derivative along $U$ of both sides of the previous equation gives the left hand side of the relative acceleration equation (\ref{eq:2}):
\begin{eqnarray}
\label{eq:secder}
\fl\quad
\nabla_{UU} Y
&=&  [Y^0 a(U)^2-a(U)\cdot (K(U)\rightcontract \vec Y)]U-[\vec Y\cdot a(U)]a(U)+\nonumber \\
\fl\quad
&&+ Y^0 \left[\dot a(U)^a E_a+\omega_{({\rm fw},U,E)}\times a(U)\right]+ \ddot Y^a\, E_a +\nonumber \\
\fl\quad
&& -2 \omega_{({\rm fw},U,E)}\times [K(U)\rightcontract \vec Y]-\omega_{({\rm fw},U,E)}\times [\omega_{({\rm fw},U,E)}\times \vec Y]+\nonumber \\
\fl\quad
&& +\dot \omega_{({\rm fw},U,E)}\times \vec Y\ ,
\end{eqnarray}
where Eq.~(\ref{conseq1}) has been taken into account and $\dot \omega_{({\rm fw},U,E)}$ stands for $\dot \omega_{({\rm fw},U,E)}^aE_a$.
It is also easy to evaluate the term $\nabla_Y a(U)$ on the right hand side of Eq.~(\ref{eq:2}); a direct calculation shows that
\begin{eqnarray}
\label{nablaYa}
\nabla_Y a(U)&=&Y^0[a(U)^2U+\dot a(U)^a E_a+\omega_{({\rm fw},U,E)}\times a(U)]+\nonumber \\
&& +Y^b\nabla(U)_b a(U)-a(U)\cdot (K(U)\rightcontract \vec Y) U\ ,
\end{eqnarray}
where $\nabla(U)a(U)=P(U)\nabla a(U)\equiv P(U)_\alpha^\mu P(U)_\beta^\nu \nabla_\nu a(U)_\mu$. As a result, from Eqs.~(\ref{eq:secder}) and (\ref{nablaYa}) we obtain
\begin{eqnarray}
\label{eq:secder2}
\fl\quad
\nabla_{UU} Y-\nabla_Y a(U)
&=& \ddot Y^a\, E_a -Y^b\nabla(U)_b a(U)-[\vec Y\cdot a(U)]a(U)+\nonumber \\
\fl\quad
&& -\omega_{({\rm fw},U,E)}\times [\omega_{({\rm fw},U,E)}\times \vec Y]\nonumber \\
\fl\quad
&& 
-2 \omega_{({\rm fw},U,E)}\times [K(U)\rightcontract \vec Y]+\dot \omega_{({\rm fw},U,E)}\times \vec Y
\nonumber \\
&=& -{\mathcal E}(U)\rightcontract \vec Y \ .
\end{eqnarray}
de Felice \cite{fdf1,fdf2,fdf3} introduced the relative strains as components of the following tensor:
\beq
\label{eq:strainsdef}
S(U)=\nabla(U) a(U)+a(U)\otimes a(U)\ ,
\eeq
namely $S(U)_{ab}=\nabla(U)_b a(U)_a+a(U)_aa(U)_b$. 
The tensor $S$ will be termed  strain tensor (actually Fermi-Walker strain tensor, see below); it depends only on the congruence $U$ and not on the chosen spatial triad $E_a$.
To make our formulas more compact we also introduce the notation
\begin{eqnarray}\fl\quad
\label{Tdef}
T_{({\rm fw},U,E)}{}^a{}_b&=& \dot C_{({\rm fw},U,E)}{}^a{}_b- [C^2_{({\rm fw},U,E)}]^a{}_b-2C_{({\rm fw},U,E)}{}^a{}_c\, K(U)^c{}_b\nonumber \\
\fl\quad
&=&\delta^a_b \omega_{({\rm fw},U,E)}^2-\omega_{({\rm fw},U,E)}^a \omega_{({\rm fw},U,E)}{}_b -\epsilon^a{}_{bf}\dot \omega_{({\rm fw},U,E)}^f\nonumber \\
\fl\quad
&& -2\epsilon^a{}_{fc}\omega_{({\rm fw},U,E)}^f K(U)^c{}_b\ , 
\end{eqnarray}
where $[C^2_{({\rm fw},U,E)}]^a{}_b=C_{({\rm fw},U,E)}{}^a{}_c\, C_{({\rm fw},U,E)}{}^c{}_b$.
The relative acceleration equation (\ref{eq:2}) (or equivalently (\ref{eq:3})) then becomes
\beq
\label{eq:secder3bis}
\ddot Y^a +{\mathcal K}_{(U,E)}{}^a{}_b  Y^b=0\ ,  
\eeq
where  
\beq
{\mathcal K}_{(U,E)}{}^a{}_b=[T_{({\rm fw},U,E)}-S(U)+{\mathcal E}(U)]^a{}_b\ .  
\eeq
Eqs.~(\ref{eq:secder3bis}) are our \lq\lq master equations" which we shall analyze 
in the following special cases:

\begin{itemize}

\item {\it Flat spacetime}: $R_{\alpha\beta\gamma\delta}=0$ so that ${\mathcal E}(U)\equiv 0$. 
In this case we have ${\mathcal K}_{(U,E)}=T_{({\rm fw},U,E)}-S(U)$.

\item $E_a$ {\it spatial triad Fermi-Walker dragged along} $U$: $\omega_{({\rm fw},U,E)}=0$.  
This implies $T_{({\rm fw},U,E)}=0$, so that ${\mathcal K}_{(U,E)}={\mathcal E}(U)-S(U)$.

\item $U$ {\it geodesic}: $a(U)\equiv 0$.  
In this case $S(U)=0$ and hence ${\mathcal K}_{(U,E)}=T_{({\rm fw},U,E)}+{\mathcal E}(U)$.

\item $U$ {\it irrotational}: $\omega(U)\equiv 0$, so that $K(U)=-\theta(U)$. 

\item $U$ {\it Born-rigid}: $\theta(U)\equiv 0$, so that $K(U)=\omega(U)$. 

\end{itemize}
Clearly we can also consider combinations of the above special cases. For example a congruence of geodesic and irrotational orbits:

\begin{itemize}

\item  $U$ {\it geodesic and irrotational}: $a(U)\equiv 0$ and $\omega(U)\equiv 0$ ($K(U)=-\theta(U)$). 
In this case $S(U)=0$, so that ${\mathcal K}_{(U,E)}=T_{({\rm fw},U,E)}+{\mathcal E}(U)$. 

\end{itemize}
Finally, the case ${\mathcal K}_{(U,E)}=0$ corresponds to $\ddot Y^a=0$, i.e. absence of relative accelerations among the particles of the congruence.

\subsection{Frame-dependent strain definition}

The derivation of the relative acceleration equation (\ref{eq:secder3bis}) for the frame components of the deviation vector has been obtained by using a generic spatial triad $E_a$ characterized by its rotation with respect to a Fermi-Walker transported triad along $U$ according to Eq.~(\ref{spat_frame}).
Actually, one can characterize the generic triad $E_a$ in different ways; for instance by its rotation with respect to other (not Fermi-Walker) geometrically meaningful frames dragged along $U$.
   
In \cite{mfg} two more derivatives of spatial vectors along $U$ have been introduced, namely

\begin{enumerate}

\item {\it Spatial co-rotating Fermi-Walker transport along $U$,} defined by the relation
\begin{eqnarray}
\fl\quad
\nabla_{\rm (cfw)}(U)E_a&=&\nabla_{\rm (fw)}(U)E_a-\omega(U)\times E_a\nonumber \\
\fl\quad
&=& [\omega_{({\rm fw},U,E)}-\omega(U)]\times E_a = C_{ ({\rm cfw},U,E)}{}^b{}_aE_b\ ,
\end{eqnarray}
whence
\beq
\nabla_{\rm (cfw)}(U)X^a=\dot X^a + C_{({\rm cfw},U,E)}{}^a{}_bX^b\ ;
\eeq

\item {\it Spatial Lie transport along $U$,} defined by the relation
\beq
\fl\quad
\nabla_{\rm (lie)}(U)E_a=\nabla_{\rm (cfw)}(U)E_a-\theta(U)^b{}_a E_b= C_{({\rm lie},U,E)}{}^b{}_aE_b\ ,
\eeq  
whence
\beq
\nabla_{\rm (lie)}(U)X^a=\dot X^a + C_{({\rm lie},U,E)}{}^a{}_bX^b\ .
\eeq

\end{enumerate}
The different structure functions are then related each other by
\beq
\fl\qquad
C_{({\rm cfw},U,E)}{}^b{}_a=C_{ ({\rm fw},U,E)}{}^b{}_a+\omega(U)^b{}_a=C_{({\rm lie},U,E)}{}^b{}_a+\theta(U)^b{}_a\ .
\eeq
Due to the fact that these three spatial \lq\lq temporal derivatives" are the only true geometrically motivated
operators, a useful notation to handle with them contemporarily has also been introduced in \cite{mfg}
\beq
\fl\quad
\{\nabla_{\rm (tem)}(U)\}_{\rm tem=fw,cfw,lie}=\{\nabla_{\rm (fw)}(U),\nabla_{\rm (cfw)}(U),\nabla_{\rm (lie)}(U)\}\ .
\eeq
Here the subscript $({\rm tem},U,E)$ means that we are referring to a $U,E_a$ tetrad and a tem=fw,cfw,lie derivative operation.
For example, the Lie transport equation (\ref{conseq2})  can be written in the equivalent forms
\beq
\fl
\nabla_{\rm (fw)}(U)\vec Y+K(U)\rightcontract \vec Y=0, \quad \nabla_{\rm (cfw)}(U)\vec Y-\theta(U)\rightcontract \vec Y=0\ , \quad
\nabla_{\rm (lie)}(U)\vec Y=0\ .
\eeq

Let us now turn to the relative acceleration equation (\ref{eq:secder3bis}); a straightforward calculation shows that
\beq
\nabla_{\rm (fw)}(U)^2\, Y^a=\ddot Y^a +T_{({\rm fw},U,E)}{}^a{}_bY^b\ ,
\eeq
hence clarifying the meaning of the spatial tensor $T_{({\rm fw},U,E)}$.
Equation (\ref{eq:secder3bis}) can then be written as
\beq
\label{strain_fw}
\nabla_{\rm (fw)}(U)^2\, \vec Y +[{\mathcal E}(U)-S(U)]\rightcontract \vec Y =0\ ,
\eeq
so that the de Felice's strain definition corresponds in the present treatment to a Fermi-Walker strain definition. 
Let us explore this circumstance in detail. 
Analogously to Eq.~(\ref{strain_fw}) one can write
\beq
\fl\quad
\nabla_{\rm (cfw)}(U)^2\, Y^a=\ddot Y^a +T_{ ({\rm cfw},U,E)}{}^a{}_bY^b\ ,\quad
\nabla_{\rm (lie)}(U)^2\, Y^a=\ddot Y^a +T_{ ({\rm lie},U,E)}{}^a{}_bY^b\ ,
\eeq
and derive the relations between $T_{ ({\rm fw},U,E)},T_{({\rm cfw},U,E)},T_{({\rm lie},U,E)}$ (all defined by the first of Eqs.~(\ref{Tdef}) using the proper functions $C_{({\rm tem},U,E)}{}^a{}_b$) so that
\begin{eqnarray}
\nabla_{\rm (fw)}(U)^2\, \vec Y&=&\nabla_{\rm (cfw)}(U)^2\, \vec Y+{\Delta T}_{({\rm cfw-fw},U,E)}\rightcontract \vec Y\nonumber \\
&=& \nabla_{\rm (lie)}(U)^2\, \vec Y+{\Delta T}_{ ({\rm lie-fw},U,E)}\rightcontract \vec Y\ ,
\end{eqnarray}
with ${\Delta T}_{({\rm cfw-fw},U,E)} =T_{({\rm fw},U,E)}-T_{ ({\rm cfw},U,E)}$, ${\Delta T}_{({\rm lie-fw},U,E)}
=T_{({\rm fw},U,E)}-T_{ ({\rm lie},U,E)}$.
Hence besides Eq.~(\ref{strain_fw}) one has the companion relations
\beq
\label{strain_cfw}
\nabla_{\rm (cfw)}(U)^2\, \vec Y +{\Delta T}_{ ({\rm cfw-fw},U,E)}\rightcontract \vec Y+ [{\mathcal E}(U)-S(U)]\rightcontract \vec Y =0\ ,
\eeq
and
\beq
\label{strain_lie}
\nabla_{\rm (lie)}(U)^2\, \vec Y +{\Delta T}_{ ({\rm lie-fw},U,E)}\rightcontract \vec Y+ [{\mathcal E}(U)-S(U)]\rightcontract \vec Y =0\ .
\eeq
The latter equation can be further simplified recalling that $\nabla_{\rm (lie)}(U)\, \vec Y=0$ and hence $\nabla_{\rm (lie)}(U)^2\, \vec Y=0$:
\beq
{\Delta T}_{({\rm lie-fw},U,E)}+{\mathcal E}(U)-S(U)=0\ .
\eeq

This analysis would allow a more involved definition of strains in terms of the kinematical properties of the frame with respect to which they are measured. For instance, one can  summarize Eqs.~(\ref{strain_fw}), (\ref{strain_cfw}) and (\ref{strain_lie}) in a single one
\beq
\label{strain_tem}
\fl\quad
\nabla_{\rm (tem)}(U)^2\, \vec Y +[{\mathcal E}(U)-S_{ ({\rm tem},U,E)}]\rightcontract \vec Y =0\ , \quad {\rm tem=fw,cfw,lie}\ ,
\eeq
and more properly identify \lq\lq {\it tem}-dependent" strains
\begin{eqnarray}
S_{({\rm fw},U,E)} &=&S(U)\ , \nonumber \\
S_{ ({\rm cfw},U,E)}&=&S_{({\rm fw},U,E)}-{\Delta T}_{({\rm cfw-fw},U,E)}\ ,\nonumber \\
S_{({\rm lie},U,E)}&=& S_{({\rm fw},U,E)}-{\Delta T}_{ ({\rm lie-fw},U,E)}\ .
\end{eqnarray}
Clearly, when the spatial frame $E_a$ undergoes a {\it tem} transport along the congruence the quantities $C_{({\rm tem},U,E)}{}^a{}_b$ and $T_{({\rm tem},U,E)}{}^a{}_b$ vanish identically and 
\beq
\nabla_{\rm (tem)}(U)\, Y^a=\dot Y^a\ ,\quad\quad \nabla_{\rm (tem)}(U)^2\, Y^a=\ddot Y^a\ , 
\eeq
so that Eq.~(\ref{strain_tem}) implies
\beq
\label{strain_tem2}
\ddot Y^a +[{\mathcal E}(U)-S_{({\rm tem},U,E)}]^a{}_b Y^b =0\ , \quad \nabla_{\rm (tem)}(U)E_a=0\ .
\eeq

The  {\it tem}-dependent analysis of strains is not just of academic interest. As a matter of fact, it is a consequence of a systematic use of spacetime splitting techniques in general relativity: the latters not only reproduce the observer point of view but also play a key role
in the assessment of a nonlocal measurement  in the exact theory of general relativity.
However, since a Fermi-Walker frame is operationally easier to set up, we shall confine our attention to the Fermi-Walker strain definition given by de Felice and coworkers \cite{fdf1,fdf2,fdf3,sem-fdf}.

\section{Minkowski spacetime}

Consider first the simplest case of flat spacetime. Let us write the Minkowski metric in standard cilyndrical coordinates $\{t, r, \phi, z\}$
\beq
\label{munkmetr}
\rmd s^2= -\rmd t^2 + \rmd r^2+ r^2\rmd \phi^2 +\rmd z^2\ , 
\eeq
and introduce the orthonormal frame
\beq
e_{\hat t}=\partial_t\ , \qquad
e_{\hat r}=\partial_r\ , \qquad
e_{\hat z}=\partial_z\ , \qquad
e_{\hat \phi}=\frac{1}{r}\partial_\phi\ .
\eeq
Consider a family of uniformly rotating particles with angular velocity $\zeta$; the four velocity $U$ of the generic particle of the congruence is then given by
\beq
\label{Uflat}
U=\Gamma (\partial_t +\zeta \partial_\phi )=\gamma (e_{\hat t} +\nu e_{\hat \phi})\ , \qquad \gamma=(1-\nu^2)^{-1/2}\ ,
\eeq
where 
\beq
\Gamma =\left( 1-r^2\zeta^2 \right)^{-1/2}\ , \qquad \zeta=\frac{\nu}{r}\ .
\eeq
A frame adapted to $U$ can be fixed as
\beq
\label{circmink}
E(U)_1=e_{\hat r}\ , \quad 
E(U)_2=\gamma (\nu e_{\hat t} + e_{\hat \phi})\ , \quad 
E(U)_3=e_{\hat z}\ . 
\eeq
We note that this frame is of Frenet-Serret type as will be introduced in Sec. 4.4.
The orbits are accelerated: $a(U)=-\gamma^2\zeta^2r\,E(U)_1$, with vanishing expansion ($\theta(U)=0$) and with vorticity vector $\omega(U)=\gamma^3\zeta \,E(U)_3$.
It is easy to show that the deviation equations (\ref{eq:secder3bis}) reduce to $\ddot Y^a=0$, since ${\mathcal K}_{(U,E)}=0$ resulting from the balancing between the strain tensor and the Fermi-Walker tensor, namely $S(U)=T_{ ({\rm fw},U,E)}$ with only nonvanishing components
\beq
S(U)_{11}=S(U)_{22}=-\gamma^4\zeta^2\ .
\eeq
The relative velocity equation (\ref{conseq2}) implies that in addition $\dot Y^a=0$, so that the spatial components of the deviation vector remain all constant along the path with respect to the frame (\ref{circmink}).

\begin{figure} 
\typeout{*** EPS figure 1}
\begin{center}
\includegraphics[scale=0.4]{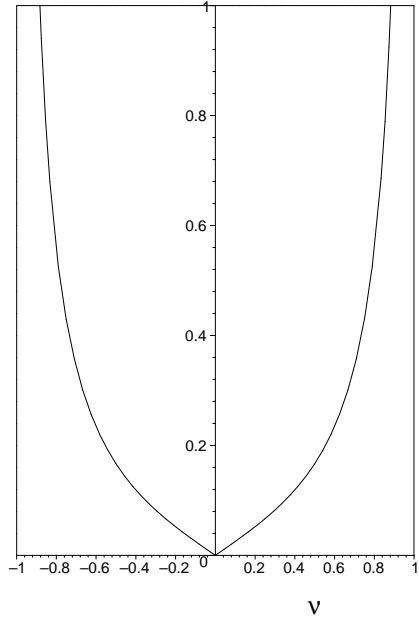}
\end{center}
\caption{Minkowski spacetime. The behaviour of the magnitude of the Fermi-Walker angular velocity $||\omega_{({\rm fw},U,E)}||$ is shown as a function of the linear velocity $\nu$ for a given circular orbit at $r=4$.
}
\label{fig:1}
\end{figure}

Rotating the spatial triad in the 2-plane $E(U)_1-E(U)_2$ by an angle $\alpha=-\gamma\zeta t=-\gamma^2\zeta\tau_U$ ($\tau_U$ denoting proper time parametrization along $U$) one obtains a Fermi-Walker triad
\begin{eqnarray}
E'(U)_1&=&\cos\alpha E(U)_1+\sin\alpha E(U)_2\ , \nonumber\\ 
E'(U)_2&=&-\sin\alpha E(U)_1+\cos\alpha E(U)_2\ , \nonumber\\ 
E'(U)_3&=&E(U)_3\ . 
\end{eqnarray}
With respect to this new triad $T_{ ({\rm fw},U,E')} =0$, so that ${\mathcal K}_{(U,E')}=-S(U)$ and the only nonvanishing components are 
\beq
{\mathcal K}_{(U,E')}{}_{11}={\mathcal K}_{(U,E')}{}_{22}=\gamma^4\zeta^2\ ,
\eeq
implying harmonic oscillations for the deviation vector components $Y^1$ and $Y^2$ with frequency $||\omega_{({\rm fw},U,E)}||=\gamma^2|\zeta|=\gamma^2 |\nu |/r$ (see Fig.~\ref{fig:1}).
This also follows from the relative velocity equations 
\beq
\dot Y'{}^1=-\gamma^2 \zeta Y'{}^2\ , \quad \dot Y'{}^2=\gamma^2 \zeta Y'{}^1\ , \quad \dot Y'{}^3=0\ .
\eeq
The corresponding solution is straightforward:
\begin{eqnarray}
Y'{}^1&=&Y'{}^1_0\cos(||\omega_{({\rm fw},U,E)}||\tau_U)- Y'{}^2_0\sin(||\omega_{({\rm fw},U,E)}||\tau_U)\ , \nonumber\\ 
Y'{}^2&=&Y'{}^2_0\cos(||\omega_{({\rm fw},U,E)}||\tau_U)+ Y'{}^1_0\sin(||\omega_{({\rm fw},U,E)}||\tau_U)\ , \nonumber\\ 
Y'{}^3&=&Y'{}^3_0\ ,
\end{eqnarray}
where $Y'{}^a_0$ are the components of the deviation vector at the starting point. 
This implies that an initially circular bunch of particles on the $Y'{}^1$-$Y'{}^2$ plane remains always circular for increasing values of the proper time.
The behaviour of the magnitude of the Fermi-Walker angular velocity as a function of the linear velocity $\nu$ is shown in Fig.~\ref{fig:1} for a fixed value of the radial coordinate.


Practically, we may think of a rigidly rotating  disk, say a merry-go-round. On such a disk one may have  bodies which are either fixed with the platform itself and comoving with it (say a \lq\lq car") or have only a fixed point and an axis parallel to the rotation axis of the disk 
about which it can rotate   freely  (say a \lq\lq montgolfier").
The connecting vector $Y$ of our previous analysis can be taken in this case as the distance between the bodies (for the \lq\lq montgolfier" this distance is meant with respect to its fixed point).
The components of this vector are constant when they are referred to a frame triad at rest with the rotating disk, but they undergo oscillations when they are referred to a frame triad fixed with  respect to the infinity (e.g., axes at rest with the \lq\lq mongolfier"). These two cases are exactly what we have   considered in the above discussion, namely the \lq\lq car" example on the merry-go-round corresponds to the first Frenet-Serret frame, while the \lq\lq mongolfier" example corresponds to the Fermi-Walker frame.

\section{Vacuum stationary axisymmetric spacetimes}
\label{kerrcase}

Consider the case of vacuum stationary axisymmetric spacetimes.
Using a coordinate system $\{t,r,\theta,\phi \}$ adapted to the spacetime symmetries, i.e. with $\partial_t$ (timelike) and $\partial_\phi$ (spacelike) a pair of commuting Killing vectors the metric can be written as
\beq
\label{metr_gen}
\rmd s^2= g_{tt} \rmd t^2 + 2 g_{t\phi} \rmd t \rmd \phi + g_{\phi\phi}\rmd \phi^2 + g_{rr}\rmd r^2 + g_{\theta\theta} \rmd \theta^2\ ,
\eeq
where all the metric coefficients depend only on $r$ and $\theta$.
Due to the spacetime symmetries it is quite natural to consider two families of observers which are described by two geometrically motivated congruences of curves:

\begin{enumerate}

\item {\it Static observers}, at rest at a given point in the spacetime; their four velocity is aligned with the Killing temporal direction
\beq
m=\frac{1}{M}\partial_t\ , \qquad M=\sqrt{-g_{tt}}\ ,
\eeq
with dual
\beq
m^\flat =-M(\rmd t-M_\phi \rmd \phi)\ , \qquad M_\phi=-g_{t\phi}/g_{tt}\ ;
\eeq
we denote by $\tau_m$ the proper time parameter along $m$ defined by $\rmd \tau_m=M\rmd t$.
We notice that $M$ and $M_\phi$ are called the lapse and shift functions for the static observers respectively, according to a terminology due to Wheeler \cite{wheeler}.
Introducing the lapse and shift notation, the spacetime metric (\ref{metr_gen}) can also be written as
\beq
\label{metr_thd}
\rmd s^2= -M^2( \rmd t-M_\phi \rmd \phi )^2 + \gamma_{\phi\phi}\rmd \phi^2 + g_{rr}\rmd r^2 + g_{\theta\theta} \rmd \theta^2\ ,
\eeq
where $\gamma_{\phi\phi}=g_{\phi\phi}+M^2M_\phi^2$; in this case
an orthonormal frame adapted  to the static observers is given by
\beq
\label{thdframe}
\fl\quad
E(m)_1= \frac{1}{\sqrt{g_{rr}}}\partial_r\ , \quad
E(m)_2=\frac{1}{\sqrt{g_{\theta\theta}}}\partial_\theta\ , \quad
E(m)_3= \frac{1}{\sqrt{\gamma_{\phi\phi}}}[\partial_\phi+M_\phi \partial_t]\ ,
\eeq
and in general is not a Fermi-Walker frame since $\omega_{ ({\rm fw},m,E)}\not =0$.
The congruence of static observers is Born-rigid: $\theta(m)=0$, but has in general  a nonzero vorticity.

In order to be at rest in the spacetime (\ref{metr_gen}) a particle must be accelerated to balance the gravitational dragging which would force it to co-rotate with the source. Evidently, if we have a bunch of particles and we want them to be at rest with respect to the infinity forming a rigid body they must be differentially accelerated. In the case of co-rotation we should consider instead the following family of observers:

\item {\it Zero Angular Momentum Observers} or ZAMO, a family of locally nonrotating observers 
with four velocity
\beq
\label{n}
\fl\qquad
n=N^{-1}(\partial_t-N^{\phi}\partial_\phi)\ , \quad N=(-g^{tt})^{-1/2}\ , \quad N^{\phi}=g_{t\phi}/g_{\phi\phi}\ ,
\eeq
with dual
\beq
n^\flat =-N\rmd t\ ;
\eeq
here $N$ and $N^{\phi}$ are the lapse and shift functions associated with the ZAMOs, respectively;
we denote by $\tau_n$ the proper time parameter along $n$ defined by $\rmd \tau_n=N\rmd t$.
The spacetime metric (\ref{metr_gen}) can then be written as
\beq
\label{metr_slic}
\rmd s^2= -N^2\rmd t^2  + g_{rr}\rmd r^2 + g_{\theta\theta} \rmd \theta^2+ g_{\phi\phi}(\rmd \phi+N^\phi \rmd t)^2\ ,
\eeq
so that a suitable orthonormal frame adapted to  the ZAMOs is now fixed by the triad
\beq
\label{zamoframe}
\fl\qquad
E(n)_1=\frac{1}{\sqrt{g_{rr}}}\partial_r\ , \quad
E(n)_2=\frac{1}{\sqrt{g_{\theta\theta}}}\partial_\theta\ , \quad
E(n)_3=\frac1{\sqrt{g_{\phi \phi }}}\partial_\phi\ .
\eeq
For the ZAMO spatial triad we use the more standard notation
\beq
e_{\hat r}=E(n)_1\ , \quad e_{\hat \theta}=E(n)_2\ , \quad e_{\hat \phi}=E(n)_3\ .
\eeq
This frame in general is not a Fermi-Walker frame since $\omega_{ ({\rm fw},n,E)}\not =0$.
The congruence of ZAMOs is irrotational: $\omega(n)=0$, but is not Born-rigid in general.

\end{enumerate}

Let now the spacetime be given by the Kerr solution. In standard Boyer-Lindquist coordinates the metric writes as
\begin{eqnarray}
\rmd s^2 &=& -\left(1-\frac{2{\mathcal M}r}{\Sigma}\right)\rmd t^2 -\frac{4a{\mathcal M} r}{\Sigma}\sin^2\theta\rmd t\rmd\phi+ \frac{\Sigma}{\Delta}\rmd r^2 +\Sigma\rmd \theta^2\nonumber\\
&&+\frac{(r^2+a^2)^2-\Delta a^2\sin^2\theta}{\Sigma}\sin^2 \theta \rmd \phi^2\ ,
\end{eqnarray}
with inverse 
\begin{eqnarray}
\fl\quad
g^{\alpha\beta}\partial_\alpha\partial_\beta &=& -\frac{1}{\Delta\Sigma}\left[ (r^2+a^2)\partial_t + a \partial_\phi\right]^2 +\frac{1}{\Sigma \sin^2 \theta} \left[ \partial_\phi+ a \sin^2 \theta  \partial_t\right]^2\nonumber\\
\fl\quad
&& +\frac{\Delta}{\Sigma}(\partial_r)^2+\frac{1}{\Sigma}(\partial_\theta)^2\ ,
\end{eqnarray}
where $\Delta=r^2-2{\mathcal M}r+a^2$ and $\Sigma=r^2+a^2\cos^2\theta$. Here $\mathcal{M}$ and $a$ are the total mass and specific angular momentum characterizing the spacetime.  The (outer) event horizon is located at $r_+={\mathcal M}+\sqrt{{\mathcal M}^2-a^2}$. 

\subsection{Kerr spacetime: Static observers}
\label{thread}

The explicit expressions for the quantities entering the relative acceleration equation  (\ref{eq:secder3bis}) and corresponding to a congruence of static observers are listed in \ref{appKthread}.
We find that the matrix ${\mathcal K}_{(m,E)}$ relative to the frame (\ref{thdframe}) is identically zero.
As a result, the system of deviation equations (\ref{eq:secder3bis}) reduces to  
\beq
\ddot Y^a=0\ .
\eeq
The first order system of Lie transport equations (\ref{conseq2}) implies that in addition 
\beq
\dot Y^a=0\ ,
\eeq
so that all the spatial components of the deviation vector remain constant along the path. This means that the system under consideration is rigid in the sense that the tidal deformation induced by the geometry is balanced by a suitable choice of added strains and triad frame.

\begin{figure} 
\typeout{*** EPS figure 2}
\begin{center}
\includegraphics[scale=0.4]{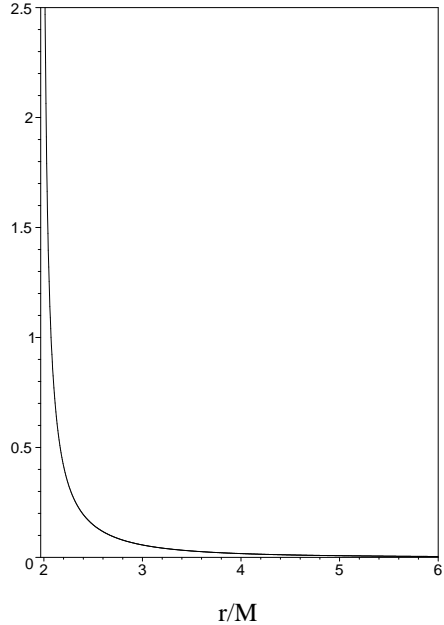}
\end{center}
\caption{Kerr spacetime, static observers. The behaviour of the magnitude of the Fermi-Walker angular velocity $||\omega_{({\rm fw},m,E)}||$ is shown as a function of $r/\mathcal{M}$, for $a/\mathcal{M}=0.5$ and $\theta=\pi/3$. Note that the static observers do not exist any longer inside the ergosphere, located at $r_{\rm erg}/\mathcal{M}\approx1.968$, where $||\omega_{({\rm fw},m,E)}||$ diverges.
}
\label{fig:2}
\end{figure}

Let us now consider how the set of deviation equations modifies when referred to a Fermi-Walker transported frame adapted to $m$.
Preliminarly, define the pair of orthogonal unit vectors $\vec \omega_{({\rm fw},m,E)},\vec \omega_{({\rm fw},m,E)}^{\perp}$, the former being aligned with the Fermi-Walker angular velocity $\omega_{({\rm fw},m,E)}$:
\begin{eqnarray}\fl\quad
\vec \omega_{({\rm fw},m,E)}&=&||\omega_{({\rm fw},m,E)}||^{-1}[\omega_{({\rm fw},m,E)}{}_{1}E(m)_1+\omega_{({\rm fw},m,E)}{}_{2}E(m)_2]\ , \nonumber\\
\fl\quad
\vec \omega_{({\rm fw},m,E)}^{\perp}&=&||\omega_{({\rm fw},m,E)}||^{-1}[-\omega_{({\rm fw},m,E)}{}_{2}E(m)_1+\omega_{({\rm fw},m,E)}{}_{1}E(m)_2]\ .
\end{eqnarray}
A Fermi-Walker transported spatial triad  is then given by
\begin{eqnarray}
\label{fwthd}\fl\quad
E'(m)_1&=&\vec \omega_{({\rm fw},m,E)}\ , \nonumber\\
\fl\quad
E'(m)_2&=&\cos(||\omega_{({\rm fw},m,E)}||\tau_m)\vec \omega_{({\rm fw},m,E)}^{\perp}-\sin(||\omega_{({\rm fw},m,E)}||\tau_m)E(m)_3\ , \nonumber\\
\fl\quad
E'(m)_3&=&\sin(||\omega_{({\rm fw},m,E)}||\tau_m)\vec \omega_{({\rm fw},m,E)}^{\perp}+\cos(||\omega_{({\rm fw},m,E)}||\tau_m)E(m)_3\ . 
\end{eqnarray}
With respect to such a frame the matrix components $T_{({\rm fw},m,E')}{}_{ab}$ are identically zero; thus the deviation matrix is given by 
${\mathcal K}_{(m,E')}={\mathcal E}(m)-S(m)$, where the components of ${\mathcal E}(m)$ and $S(m)$ are now evaluated with respect to the new frame. 
The matrix ${\mathcal K}_{(m,E')}$ results to be diagonal, with nonvanishing components 
${\mathcal K}_{(m,E')}{}_{22}=||\omega_{({\rm fw},m,E)}||^2={\mathcal K}_{(m,E')}{}_{33}$, 
leading to oscillating behaviours of the components $Y'{}^2$ and $Y'{}^3$ of the deviation vector with the same frequency $||\omega_{({\rm fw},m,E)}||$.
The component $Y'{}^1$, directed along the Fermi-Walker angular velocity, remains instead constant along the path.
This follows from the relative velocity equations (\ref{conseq2}), namely
\beq
\dot Y'{}^1=0\ , \quad \dot Y'{}^2=- ||\omega_{({\rm fw},m,E)}||  Y'{}^3\ , \quad \dot Y'{}^3= ||\omega_{({\rm fw},m,E)}||  Y'{}^2\ ,
\eeq
which can be easily integrated
\begin{eqnarray}
Y'{}^1&=&Y'{}^1_0\ , \nonumber\\ 
Y'{}^2&=&Y'{}^2_0\cos(||\omega_{({\rm fw},m,E)}||\tau_m)-  Y'{}^3_0\sin(||\omega_{({\rm fw},m,E)}||\tau_m)\ , \nonumber\\ 
Y'{}^3&=&Y'{}^3_0\cos(||\omega_{({\rm fw},m,E)}||\tau_m)+Y'{}^2_0\sin(||\omega_{({\rm fw},m,E)}||\tau_m)\ ,
\end{eqnarray}
where $Y'{}^a_0$ denote the components of the deviation vector at the starting point. 
It implies that an initially circular bunch of particles on the $Y'{}^2$-$Y'{}^3$ plane does not change its shape as the proper time varies.
The behaviour of the magnitude of the Fermi-Walker angular velocity as a function of the radial coordinate is shown in Fig.~\ref{fig:2}.

In general the oscillating variation of $Y'{}^2$ and $Y'{}^3$ is  due  to the relative motion with respect to the chosen Fermi-Walker frame. In our case the $Y'{}^1$ component being stably aligned with the $\hat 1$-leg of the triad
locally identifies the Fermi-Walker rotation axis.
This particular behaviour suggests a possible experiment to measure the gravitational dragging.


\subsection{Kerr spacetime: ZAMOs}

The explicit expressions for the quantities entering the relative acceleration equation (\ref{eq:secder3bis}) and corresponding to a congruence of ZAMOs are listed in \ref{appKZAMO}.
We find that the matrix ${\mathcal K}_{(n,E)}$ is identically zero.
As a result, the system of deviation equations (\ref{eq:secder3bis}) trivially reduces to  
\beq
\label{eq:417}
\ddot Y^a=0\ .
\eeq
The first order system of the Lie transport equations (\ref{conseq2}) leads to  
\beq
\label{eq1ordzamo}
\dot Y^1=0\ , \quad \dot Y^2=0\ , \quad \dot Y^3=2[\omega_{({\rm fw},n,E)}{}_2Y^1_0-\omega_{({\rm fw},n,E)}{}_1Y^2_0]\ ,
\eeq
where the components $Y^1_0$ and $Y^2_0$ of the deviation vector, calculated at the initial time, remain constant along the path.
The solution for the component $Y^3$ can thus be written as 
\beq
Y^3=\dot Y^3_0\tau_n+Y^3_0\ ,
\eeq
where the quantity $\dot Y^3_0$ is given by the right hand side of the last equation (\ref{eq1ordzamo}) evaluated at the starting point, implying linear deviation along the direction $E(n)_3$ as a function of the proper time.
This is clearly related to the fact that close-by particles moving along circular orbits at different values of the radial coordinate $r$ have different angular velocity, the latter being that of the gravitational dragging. Evidently, since $Y^3$ changes uniformly the system of particles is not internally accelerated as from Eq.~(\ref{eq:417}); however, as stated above, it is not a rigid frame.

\begin{figure} 
\typeout{*** EPS figure 3}
\begin{center}
$\begin{array}{cccc}
\includegraphics[scale=0.42]{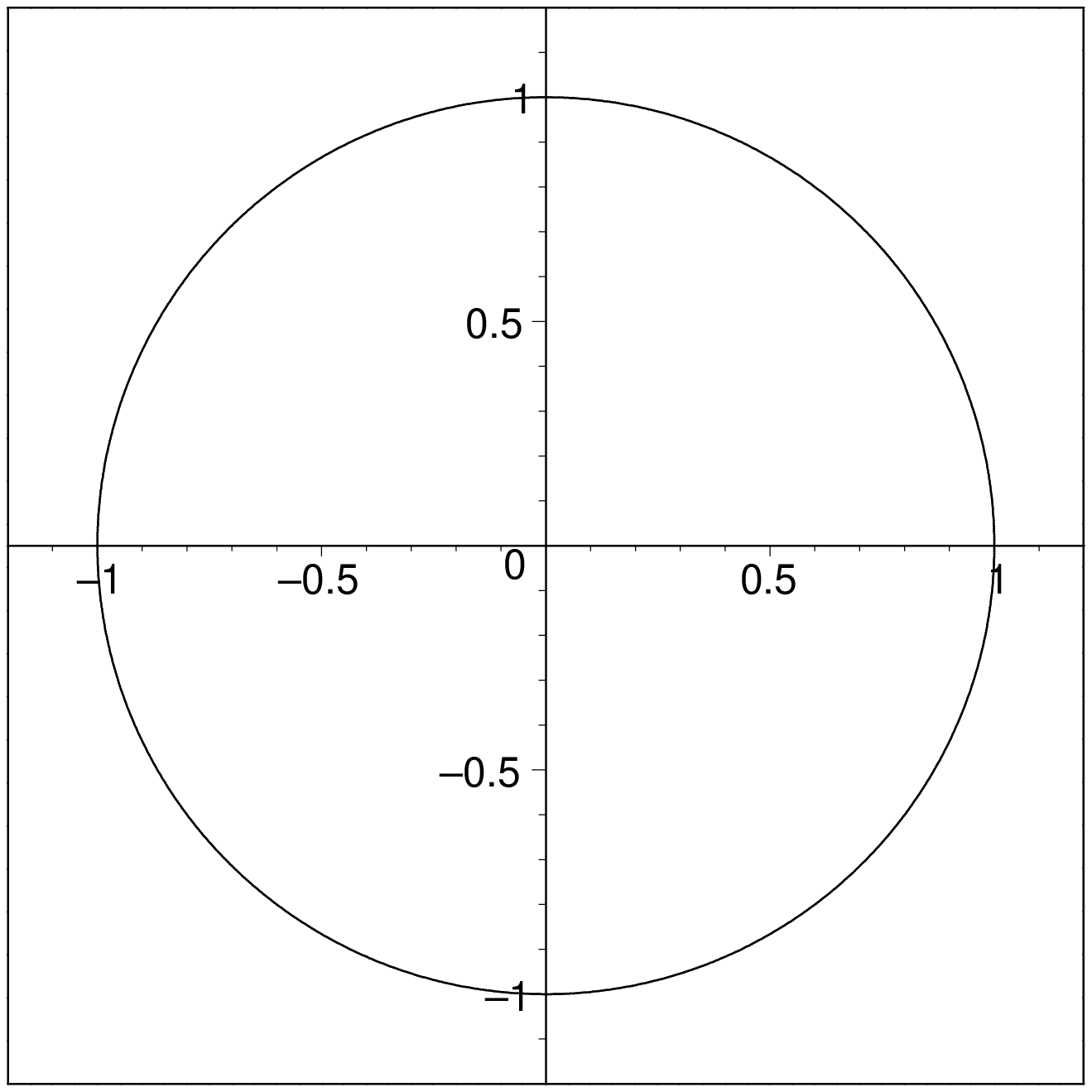}&\qquad
\includegraphics[scale=0.42]{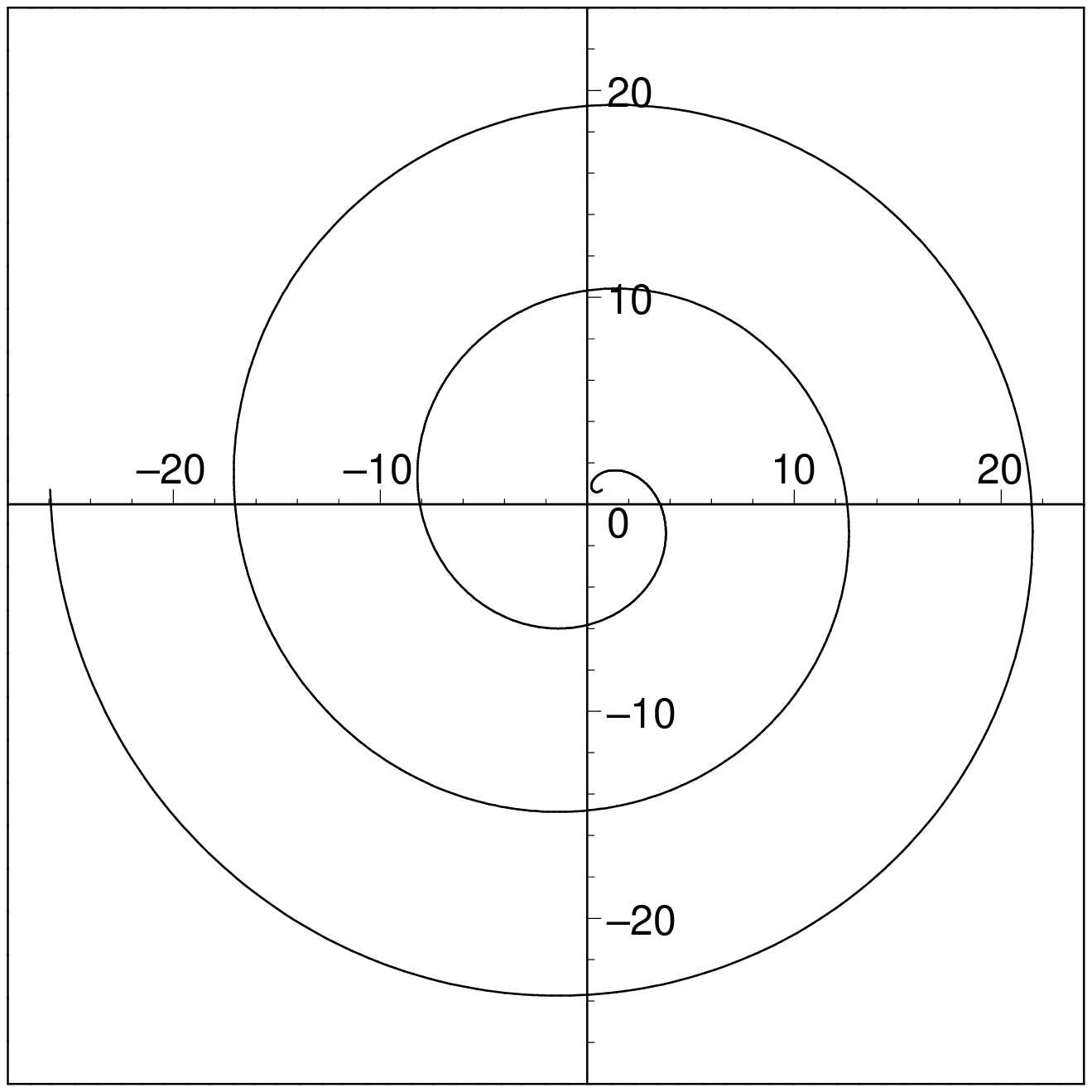}&\\[.2cm]
\mbox{(a)} &\qquad \mbox{(b)}
\end{array}
$\\
\end{center}
\caption{Kerr spacetime, ZAMOs. The components $Y'{}^3$ vs. $Y'{}^2$ of the deviation vector are shown for increasing values of the proper time $\tau_n\in[0,3T_n]$, where $T_n=2\pi/||\omega_{({\rm fw},n,E)}||$, for the choice of parameters $a/\mathcal{M}=.5$, $r/\mathcal{M}=10$ and $\theta=\pi/3$. 
In Fig.~(a) the initial conditions are set to $Y'{}^3_0=0$ and $Y'{}^2_0=1$, implying oscillating behaviour for both components as from Eq.~(\ref{solY23zamopr}). 
A nonvanishing value of $Y'{}^3_0$ gives rise instead to a spiraling behaviour which is shown in Fig.~(b) for the choice $Y'{}^3_0=1$ and $Y'{}^2_0=0$. 
}  
\label{fig:3}
\end{figure}

\begin{figure} 
\typeout{*** EPS figure 4}
\begin{center}
$\begin{array}{cccc}
\includegraphics[scale=0.4]{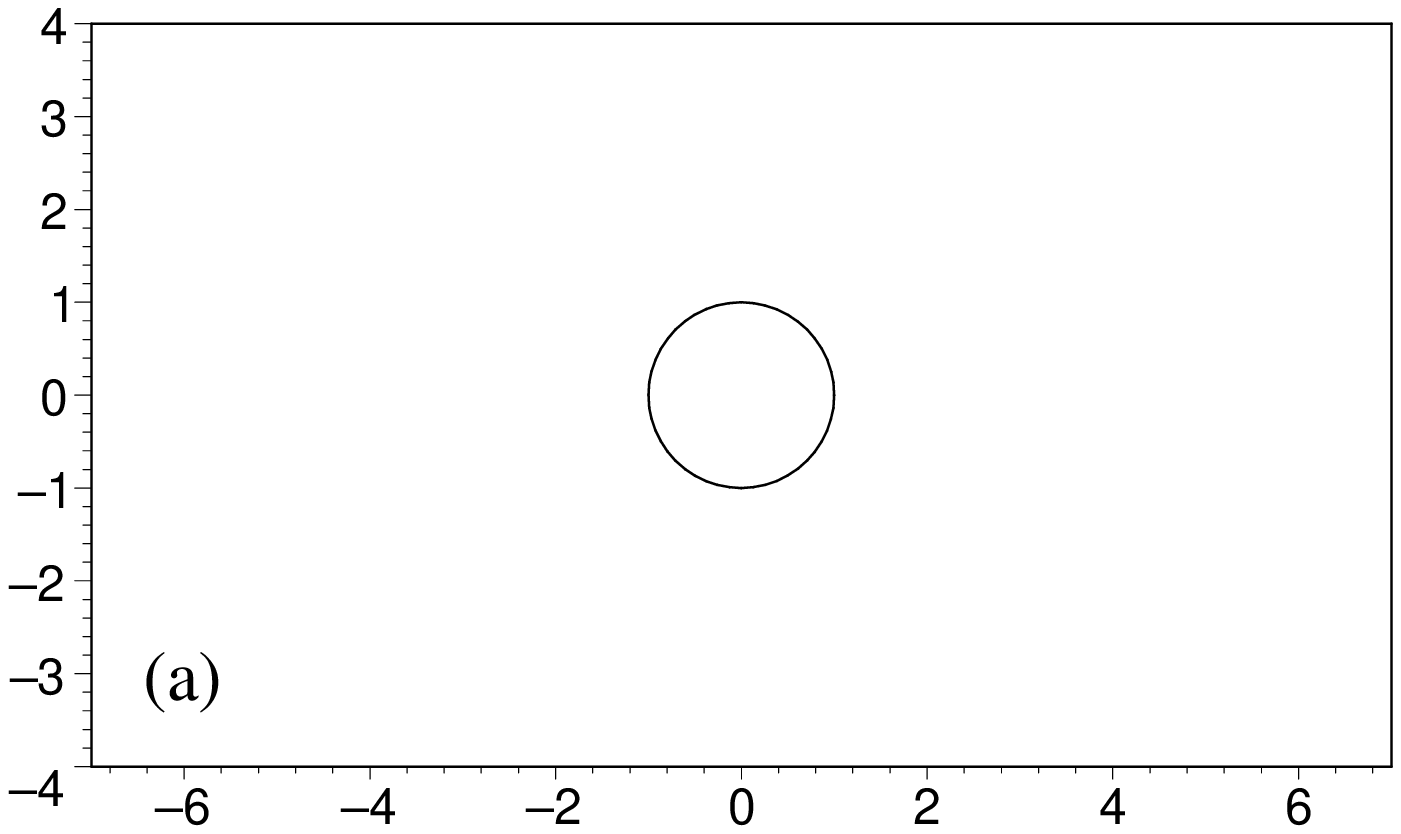}&\quad
\includegraphics[scale=0.4]{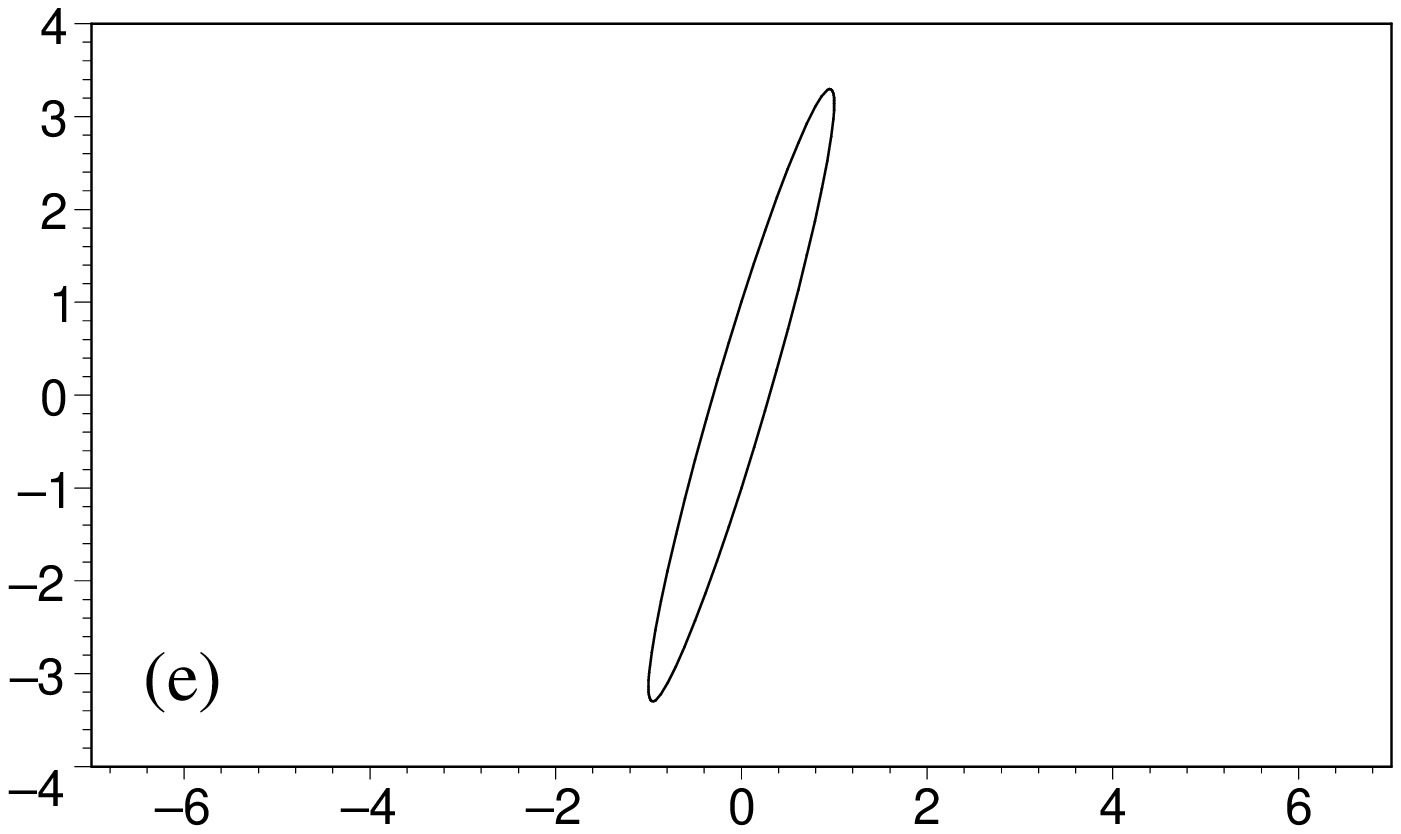}&\\[.2cm]
\includegraphics[scale=0.4]{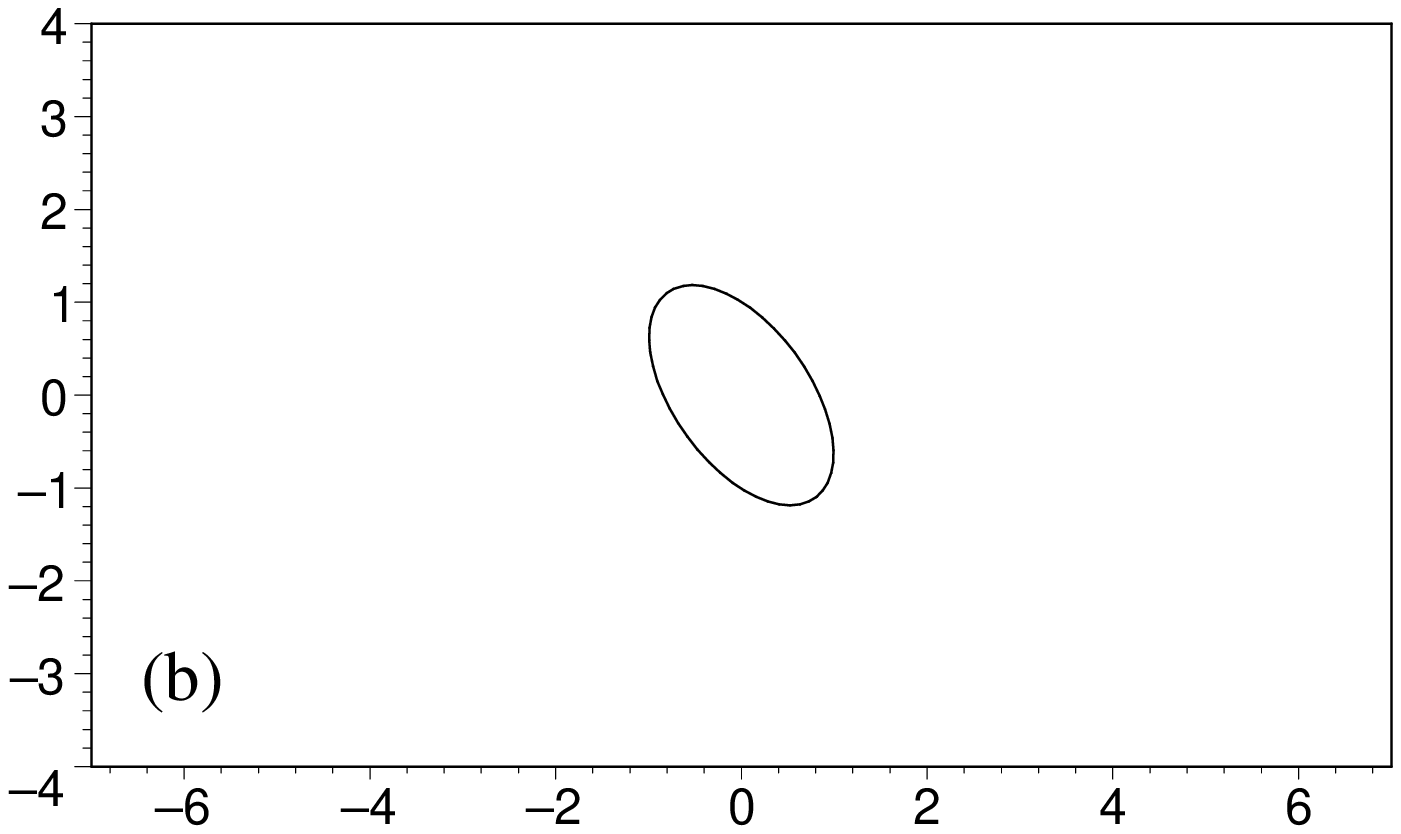}&\quad
\includegraphics[scale=0.4]{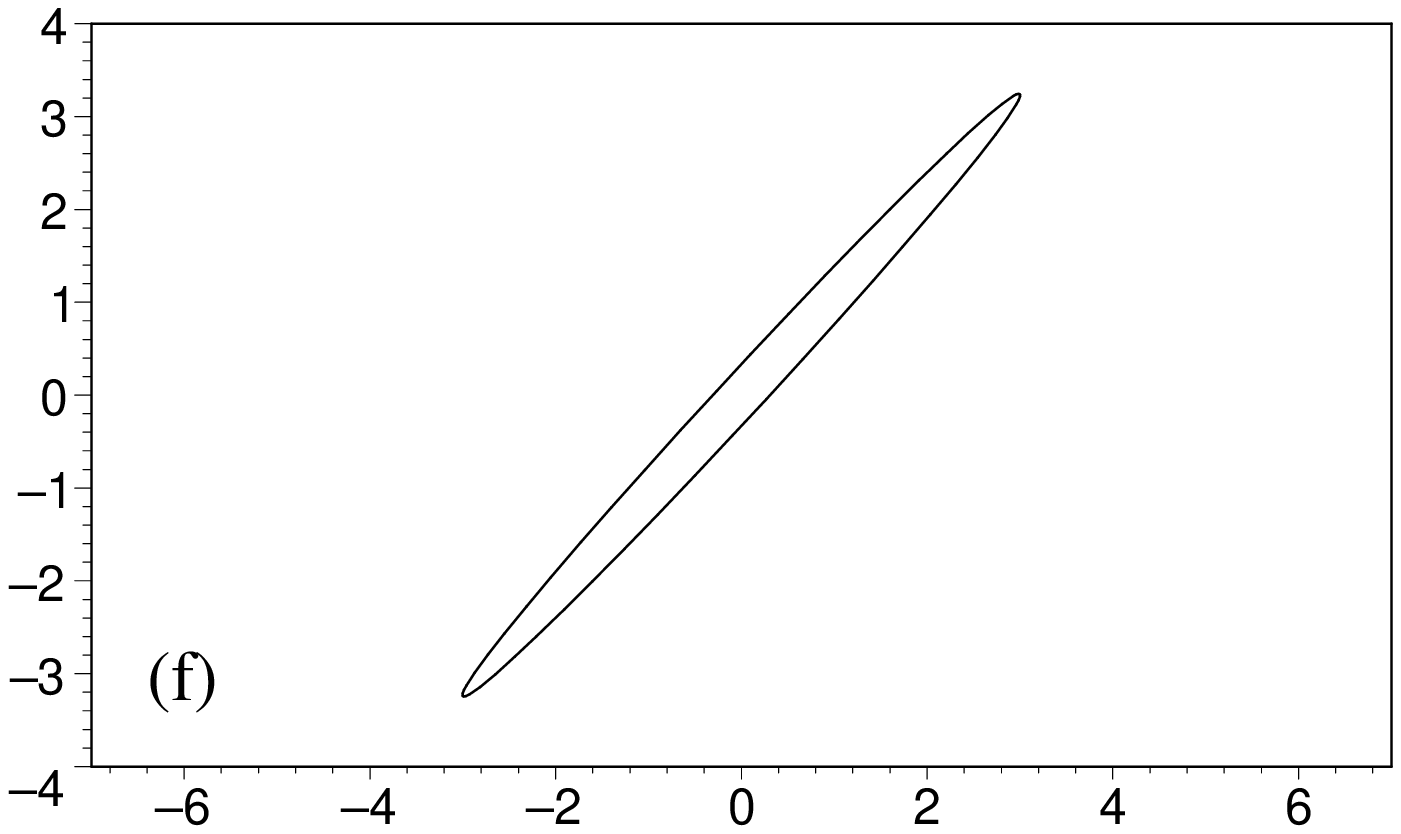}&\\[.2cm]
\includegraphics[scale=0.4]{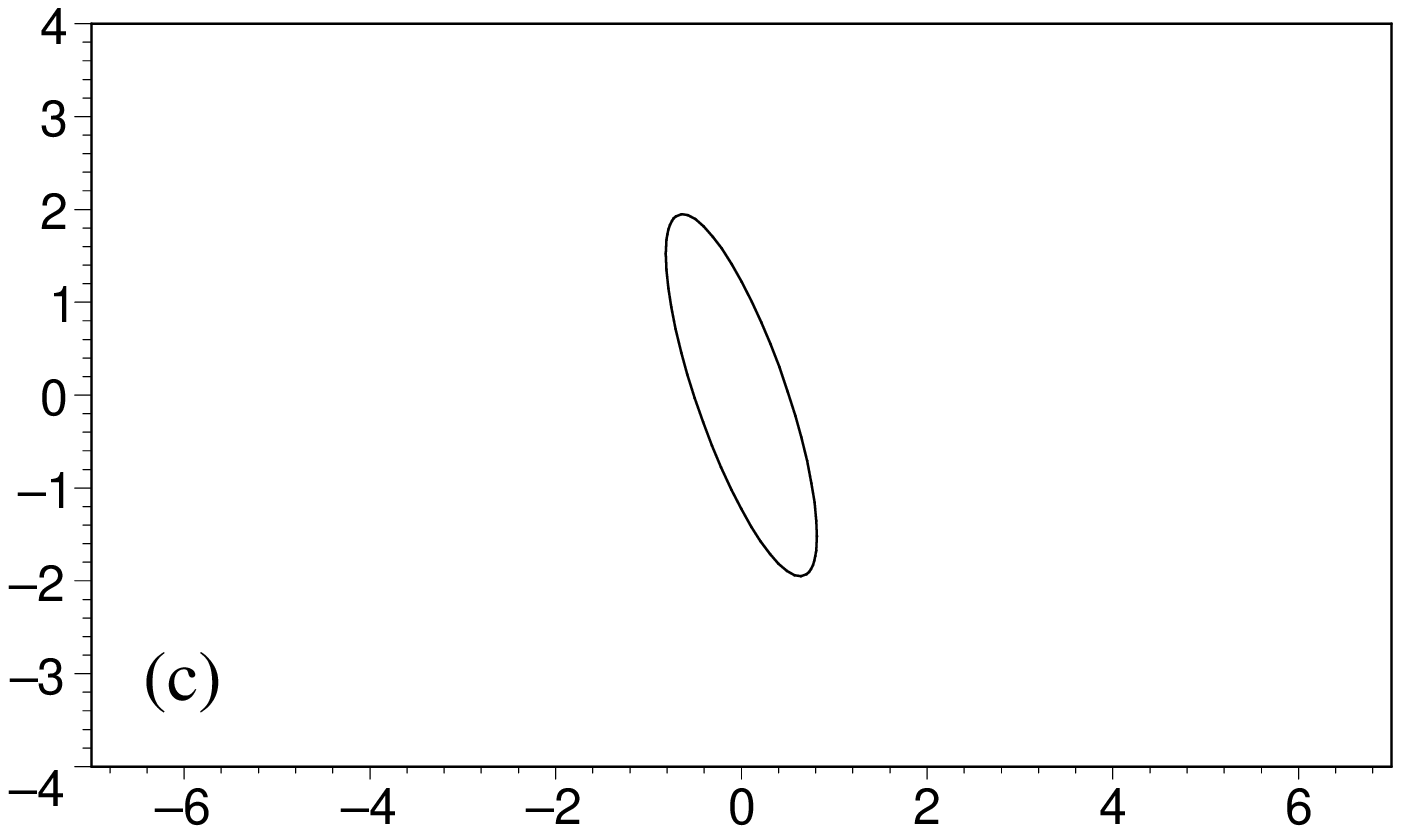}&\quad
\includegraphics[scale=0.4]{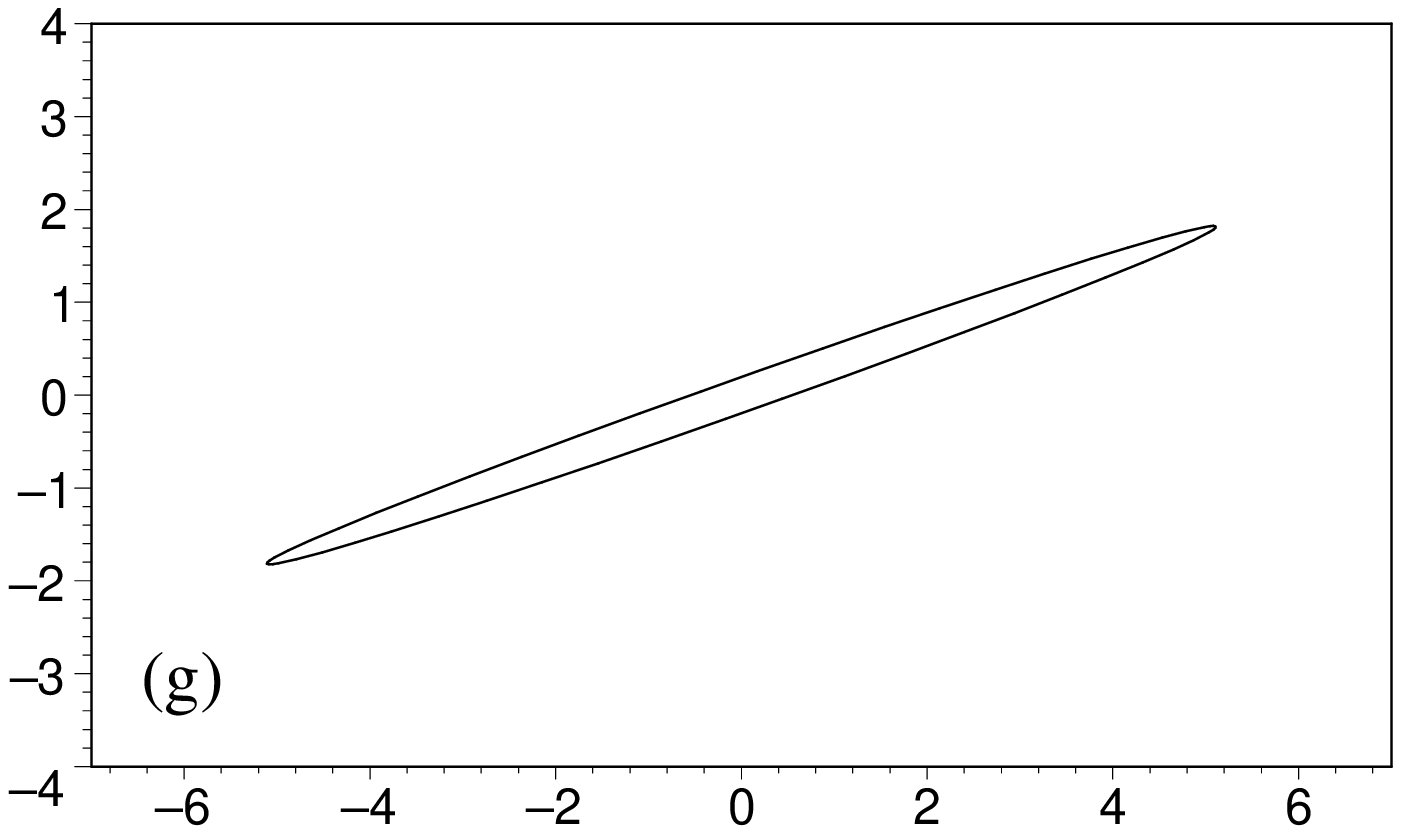}\\[.2cm]
\includegraphics[scale=0.4]{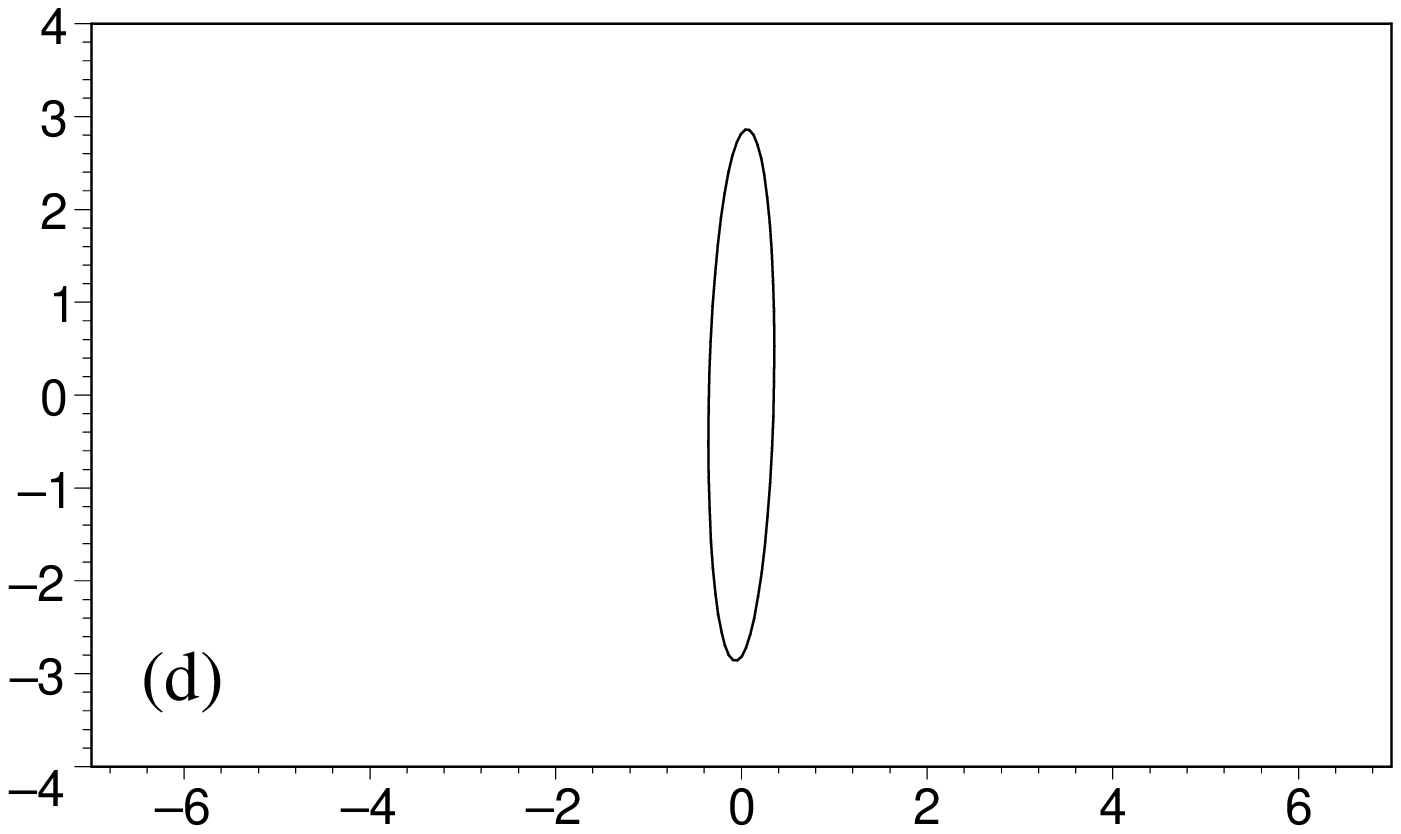}&\quad
\includegraphics[scale=0.4]{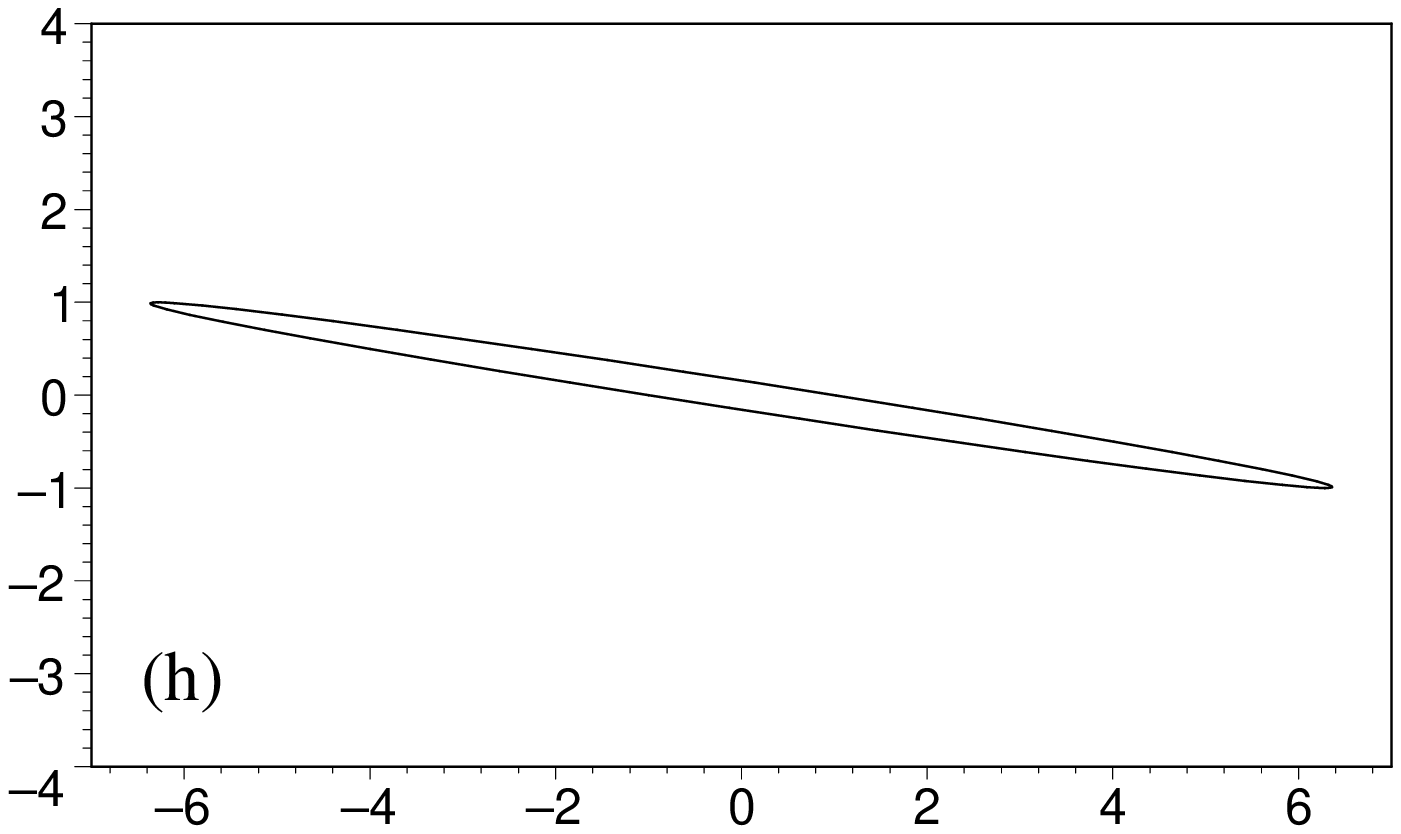}
\end{array}
$\\
\end{center}
\caption{Kerr spacetime, ZAMOs. An initially circular bunch of particles on the $Y'{}^2$-$Y'{}^3$ plane (see Fig.~(a)) is squeezed according to Fig.~\ref{fig:3} (b). The choice of parameters is the same as in Fig.~\ref{fig:5}: $a/\mathcal{M}=.5$, $r/\mathcal{M}=10$ and $\theta=\pi/3$.  
The curves depicted in Figs.~(b) to (h) correspond to increasing values of the the proper time $\tau_n/T_n=[1/20,1/8,1/5,1/4,1/3,1/2.4,1/2]$ respectively, where $T_n=2\pi/||\omega_{({\rm fw},n,E)}||$.  
}  
\label{fig:4}
\end{figure}

\begin{figure} 
\typeout{*** EPS figure 5}
\begin{center}
\includegraphics[scale=0.4]{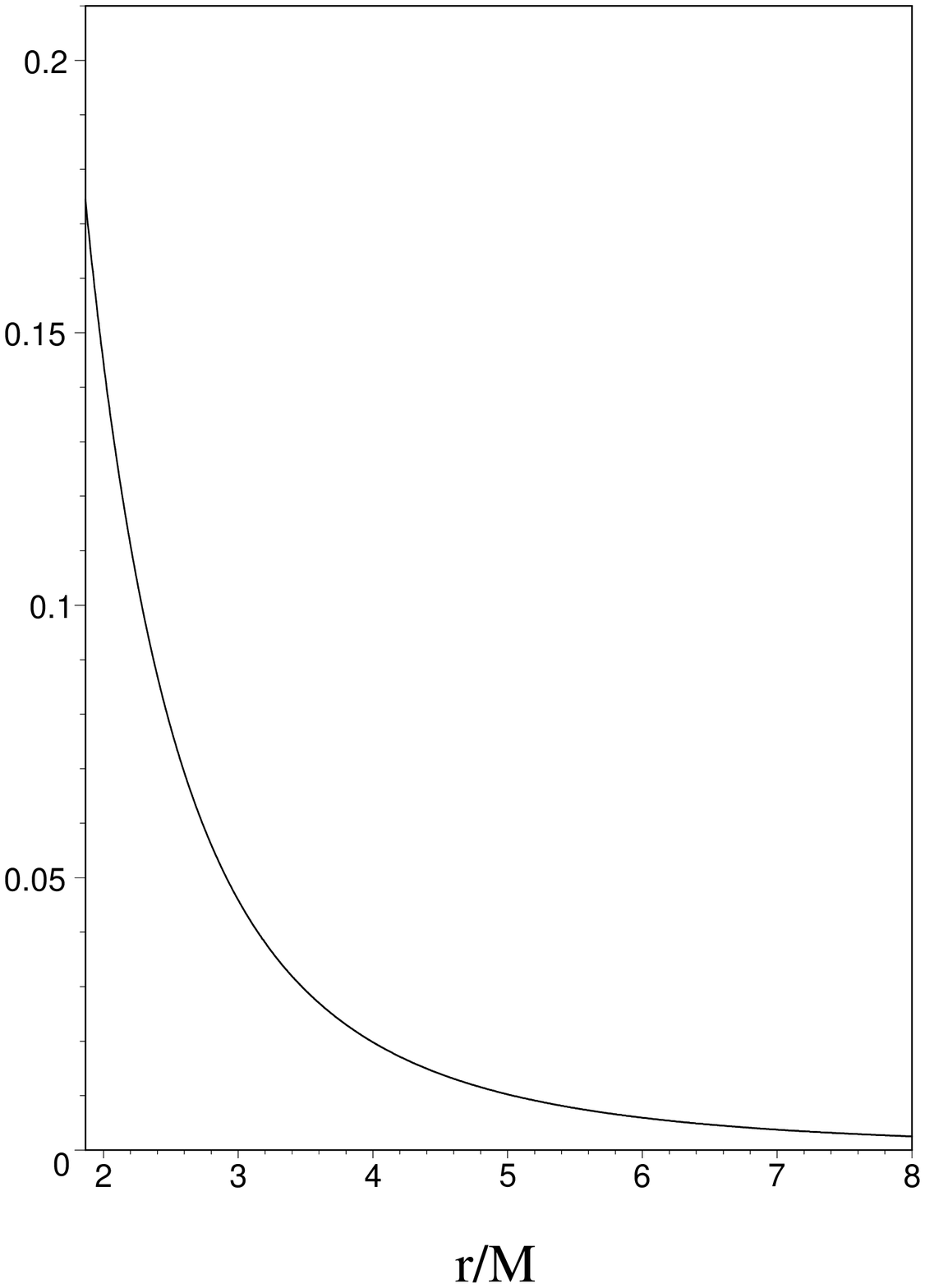}
\end{center}
\caption{Kerr spacetime, ZAMOs. The behaviour of the magnitude of the Fermi-Walker angular velocity $||\omega_{({\rm fw},n,E)}||$ is shown as a function of $r/\mathcal{M}$, for $a/\mathcal{M}=0.5$ and $\theta=\pi/3$. The horizon is located at $r_+/\mathcal{M}\approx 1.866$, where $||\omega_{({\rm fw},n)}(n)||\mathcal{M}\approx0.174$.
}
\label{fig:5}
\end{figure}

As in the case of a static observer it is interesting to write the set of deviation equations with respect to a Fermi-Walker transported frame adapted to $n$.
As before, let us introduce the pair of orthogonal unit vectors $\vec \omega_{({\rm fw},n,E)}{},\vec \omega_{({\rm fw},n,E)}{}^{\perp}$, the former being aligned with the Fermi-Walker angular velocity $\omega_{({\rm fw},n,E)}{}$: 
\begin{eqnarray}
\fl\quad
\vec \omega_{({\rm fw},n,E)}{}&=&||\omega_{({\rm fw},n,E)}||{}^{-1}[\omega_{({\rm fw},n,E)}{}_{1}E(n)_1+\omega_{({\rm fw},n,E)}{}_{2}E(n)_2]\ , \nonumber\\
\fl\quad
\vec \omega_{({\rm fw},n,E)}{}^{\perp}&=&||\omega_{({\rm fw},n,E)}||{}^{-1}[-\omega_{({\rm fw},n,E)}{}_{2}E(n)_1+\omega_{({\rm fw},n,E)}{}_{1}E(n)_2]\ ;
\end{eqnarray}
a Fermi-Walker transported spatial triad  is then given by
\begin{eqnarray}
\label{fwzamo}
\fl\quad
E'(n)_1&=&\vec \omega_{({\rm fw},n,E)}{}\ , \nonumber\\
\fl\quad
E'(n)_2&=&\cos(||\omega_{({\rm fw},n,E)}||{}\tau_n)\vec \omega_{({\rm fw},n,E)}{}^{\perp}-\sin(||\omega_{({\rm fw},n,E)}||{}\tau_m)E(n)_3\ , \nonumber\\
\fl\quad
E'(n)_3&=&\sin(||\omega_{({\rm fw},n,E)}||{}\tau_n)\vec \omega_{({\rm fw},n,E)}{}^{\perp}+\cos(||\omega_{({\rm fw},n,E)}||{}\tau_m)E(n)_3\ . 
\end{eqnarray}
The only nonvanishing components of the deviation matrix ${\mathcal K}_{(n,E')}{}={\mathcal E}(n)-S(n)$ with respect to the new triad (\ref{fwzamo}) turn out to be
\begin{eqnarray}
\fl\quad
{\mathcal K}_{(n,E')}{}_{22}&=&||\omega_{({\rm fw},n,E)}||^2[1-4\cos^2(||\omega_{({\rm fw},n,E)}||{}\tau_n)]\ , \nonumber\\
\fl\quad
{\mathcal K}_{(n,E')}{}_{23}&=&-4||\omega_{({\rm fw},n,E)}||^2\sin(||\omega_{({\rm fw},n,E)}||{}\tau_n)\cos(||\omega_{({\rm fw},n,E)}||{}\tau_n)\ , \nonumber\\
\fl\quad
{\mathcal K}_{(n,E')}{}_{33}&=&-||\omega_{({\rm fw},n,E)}||^2[3-4\cos^2(||\omega_{({\rm fw},n,E)}||{}\tau_n)]\ .
\end{eqnarray}
The relative velocity equations (\ref{conseq2}) are given by
\begin{eqnarray}\fl\quad
\dot Y'{}^1&=& 0\ , \nonumber\\
\fl\quad
\dot Y'{}^2&=& -||\omega_{({\rm fw},n,E)}||\, [\sin(2||\omega_{({\rm fw},n,E)}||\tau_n)Y'{}^2+\cos(2||\omega_{({\rm fw},n,E)}||\tau_n)Y'{}^3]\ , \nonumber\\
\fl\quad
\dot Y'{}^3&=& -||\omega_{({\rm fw},n,E)}||\, [\cos(2||\omega_{({\rm fw},n,E)}||\tau_n)Y'{}^2-\sin(2||\omega_{({\rm fw},n,E)}||\tau_n)Y'{}^3]\ ,
\end{eqnarray}
and can be easily integrated
\begin{eqnarray}
\label{solY23zamopr}
Y'{}^1&=&Y'{}^1_0\ , \nonumber\\ 
Y'{}^2&=&Y'{}^2_0\cos(||\omega_{({\rm fw},n,E)}||\tau_n)+  Y'{}^3_0\sin(||\omega_{({\rm fw},n,E)}||\tau_n)\nonumber\\ 
&&-2||\omega_{({\rm fw},n,E)}||\tau_n\cos(||\omega_{({\rm fw},n,E)}||\tau_n)Y'{}^3_0\ , \nonumber\\ 
Y'{}^3&=&-Y'{}^2_0\sin(||\omega_{({\rm fw},n,E)}||\tau_n)+  Y'{}^3_0\cos(||\omega_{({\rm fw},n,E)}||\tau_n)\nonumber\\ 
&&+2||\omega_{({\rm fw},n,E)}||\tau_n\sin(||\omega_{({\rm fw},n,E)}||\tau_n)Y'{}^3_0\ ,
\end{eqnarray}
where $Y'{}^a_0$ are the components of the deviation vector at the starting point. 
The behaviour of the components $Y'{}^3$ vs. $Y'{}^2$ for increasing values of the proper time $\tau_n$ is shown in Fig.~\ref{fig:3} for different setting of the initial conditions. 
The resulting squeezing of an initially circular bunch of particles on the $Y'{}^2$-$Y'{}^3$ plane is depicted in Fig.~\ref{fig:4}.
Figure \ref{fig:5} shows instead the behaviour of the magnitude of the Fermi-Walker angular velocity as a function of the radial coordinate.


\subsection{Kerr spacetime: Painlev\'e-Gullstrand observers}

In the Kerr spacetime it is also interesting to study the Painlev\'e-Gullstrand geodesic and irrotational family of orbits \cite{doran}.
The associated  four velocity 1-form, denoted by $\mathcal{N}^\flat$, is given by
\beq
\label{PG4vel}
\mathcal{N}^\flat=-\rmd t -\frac{\sqrt{2\mathcal{M}r(r^2+a^2)}}{\Delta}\, \rmd r\ . 
\eeq
It is easy to see that
\beq
\label{calN}
\mathcal{N}=\gamma_{(\mathcal{N},n)}[n+\nu(\mathcal{N},n) e_{\hat r}]\ ,
\eeq
so that the Painlev\'e-Gullstrand geodesics observers move radially with respect to the ZAMOs with a 
relative speed 
\beq
\fl\quad
\nu(\mathcal{N},n)=-\sqrt{\frac{2\mathcal{M}r(r^2+a^2)}{(r^2+a^2)^2-a^2\Delta\sin^2\theta}}=-\sqrt{1-N^2}\ , \quad
\gamma_{(\mathcal{N},n)}=\frac{1}{N}\ .
\eeq

We denote by $\tau_\mathcal{N}$ the proper time parameter along $\mathcal{N}$ defined by $\rmd \tau_\mathcal{N}=N^2\rmd t$.
A  frame adapted to the Painlev\'e-Gullstrand observers can be fixed with the triad
\beq
\label{framePG}
\fl\quad
E(\mathcal{N})_1=\gamma_{(\mathcal{N},n)}[\nu(\mathcal{N},n) n+E(n)_1]\ , \quad 
E(\mathcal{N})_2= E(n)_2\ , \quad
E(\mathcal{N})_3= E(n)_3\ .
\eeq
The most relevant quantities associated with the Painlev\'e-Gullstrand observers are listed in \ref{appKPG}.
The spatial components of the deviation vector are then obtained solving the coupled system of second order differential equations (\ref{eq:secder3bis}), where the nonvanishing components of ${\mathcal K}_{{(\mathcal N},E)}$ are given by Eq.~(\ref{KabPG}). They are all functions of the radial coordinate 
$r$ which in the case of Painlev\'e-Gullstrand observers is depending on the proper time $\tau_\mathcal{N}$. Therefore, the integration of the equations can be performed numerically.
The corresponding first order system of Lie transport equations (\ref{conseq2}) writes as 
\begin{eqnarray}
\label{eq1ordPG}
\dot Y^1&=&\theta(\mathcal{N})_{11}Y^1+2\omega_{({\rm fw},\mathcal{N},E)}{}_3Y^2\ , \nonumber\\ 
\dot Y^2&=&\theta(\mathcal{N})_{22}Y^2\ , \nonumber\\ 
\dot Y^3&=&\theta(\mathcal{N})_{33}Y^3+2[\omega_{({\rm fw},\mathcal{N},E)}{}_2Y^1-\omega_{({\rm fw},\mathcal{N},E)}{}_1Y^2]\ .
\end{eqnarray}

\begin{figure} 
\typeout{*** EPS figure 6}
\begin{center}
\includegraphics[scale=0.4]{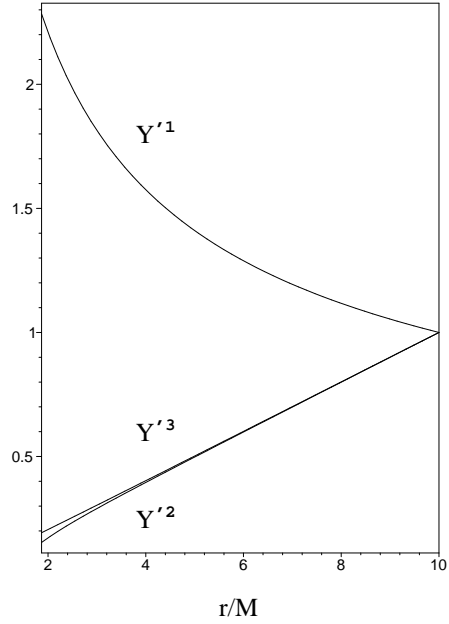}
\end{center}
\caption{Kerr spacetime, Painlev\'e-Gullstrand observers. The behaviours of the components $Y'{}^a$ of the deviation vector are shown as functions of $r/\mathcal{M}$  for the choice of parameters $a/\mathcal{M}=.5$ and $\theta=\pi/3$ (the horizon is thus located at $r_+/\mathcal{M}\approx1.866$). 
The initial conditions are set to $Y'{}^a(r_0)=1$ and $r_0/\mathcal{M}=10$.
We see that $Y'{}^1$ increases for decreasing $r$, whereas the deviations along the angular directions are both decreasing.
}
\label{fig:6}
\end{figure}

As above, one can also write the set of deviation equations with respect to a Fermi-Walker transported frame adapted to ${\mathcal N}$.
A straigthforward calculation shows that a Fermi-Walker transported spatial triad is given by
\begin{eqnarray}
\label{fwPG}
\fl\quad
E'({\mathcal N})_1&=&-\sin\beta [-\sin\alpha E(\mathcal {N})_1+\cos\alpha E(\mathcal {N})_3]-\cos\beta E(\mathcal {N})_2\ , \nonumber\\
\fl\quad
E'({\mathcal N})_2&=&-\cos\beta [-\sin\alpha E(\mathcal {N})_1+\cos\alpha E(\mathcal {N})_3]+\sin\beta E(\mathcal {N})_2\ , \nonumber\\
\fl\quad
E'({\mathcal N})_3&=&\cos\alpha E(\mathcal {N})_1+\sin\alpha E(\mathcal {N})_3\ , 
\end{eqnarray}
where 
\beq\fl\quad
\alpha=\arctan\left(\frac{\sqrt{\Sigma(r^2+a^2)}}{\sqrt{2\mathcal{M}r}a\sin\theta}\right)\ , \quad 
\beta=\theta+\arctan\left(\cot\theta\frac{\sqrt{r^2+a^2}}{r}\right)\ .
\eeq
The only nonvanishing components of the deviation matrix ${\mathcal K}_{({\mathcal N},E')}$ coincide with ${\mathcal E}'({\mathcal N})$, whose frame components are given by Eq.~(\ref{KabPGfw}).
The relative velocity equations (\ref{conseq2}) turn out to be
\begin{eqnarray}
\label{eq1ordPGfw}
\dot Y'{}^1&=& \theta'({\mathcal N})^1{}_1Y'{}^1+\theta'({\mathcal N})^1{}_2Y'{}^2\ , \nonumber\\
\dot Y'{}^2&=& \theta'({\mathcal N})^2{}_1Y'{}^1+\theta'({\mathcal N})^2{}_2Y'{}^2\ , \nonumber\\
\dot Y'{}^3&=& \theta'({\mathcal N})^3{}_3Y'{}^3\ ,
\end{eqnarray} 
where $\theta'({\mathcal N})^a{}_b$ are the components of $\theta({\mathcal N})$ with respect to the Fermi-Walker triad.
This system of first order deviation equations can be integrated numerically 
treating as integration variable the radial coordinate $r$.
The components $Y'{}^a$ in fact can be made explicitly 
depending on the radial coordinate $r$ instead of the proper time using the relation $\rmd r/\rmd\tau_\mathcal{N}=\gamma_{(\mathcal{N},n)}\nu_{(\mathcal{N},n)}/\sqrt{g_{rr}}$. 
Their behaviours as functions of $r$ are shown in Fig.~\ref{fig:6}.


\begin{figure} 
\typeout{*** EPS figure 7}
\begin{center}
$\begin{array}{cccc}
\includegraphics[scale=0.4]{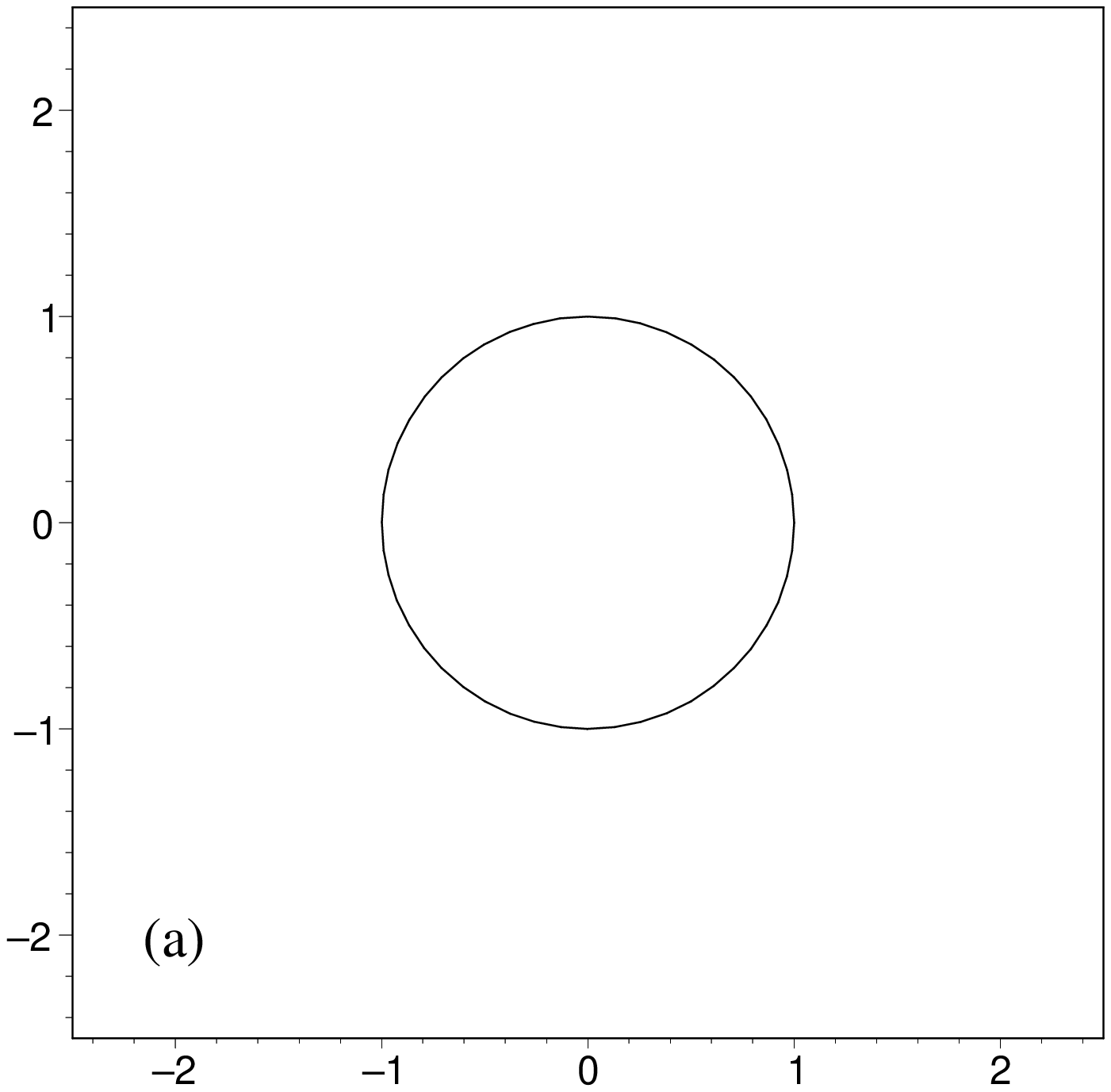}&\quad
\includegraphics[scale=0.4]{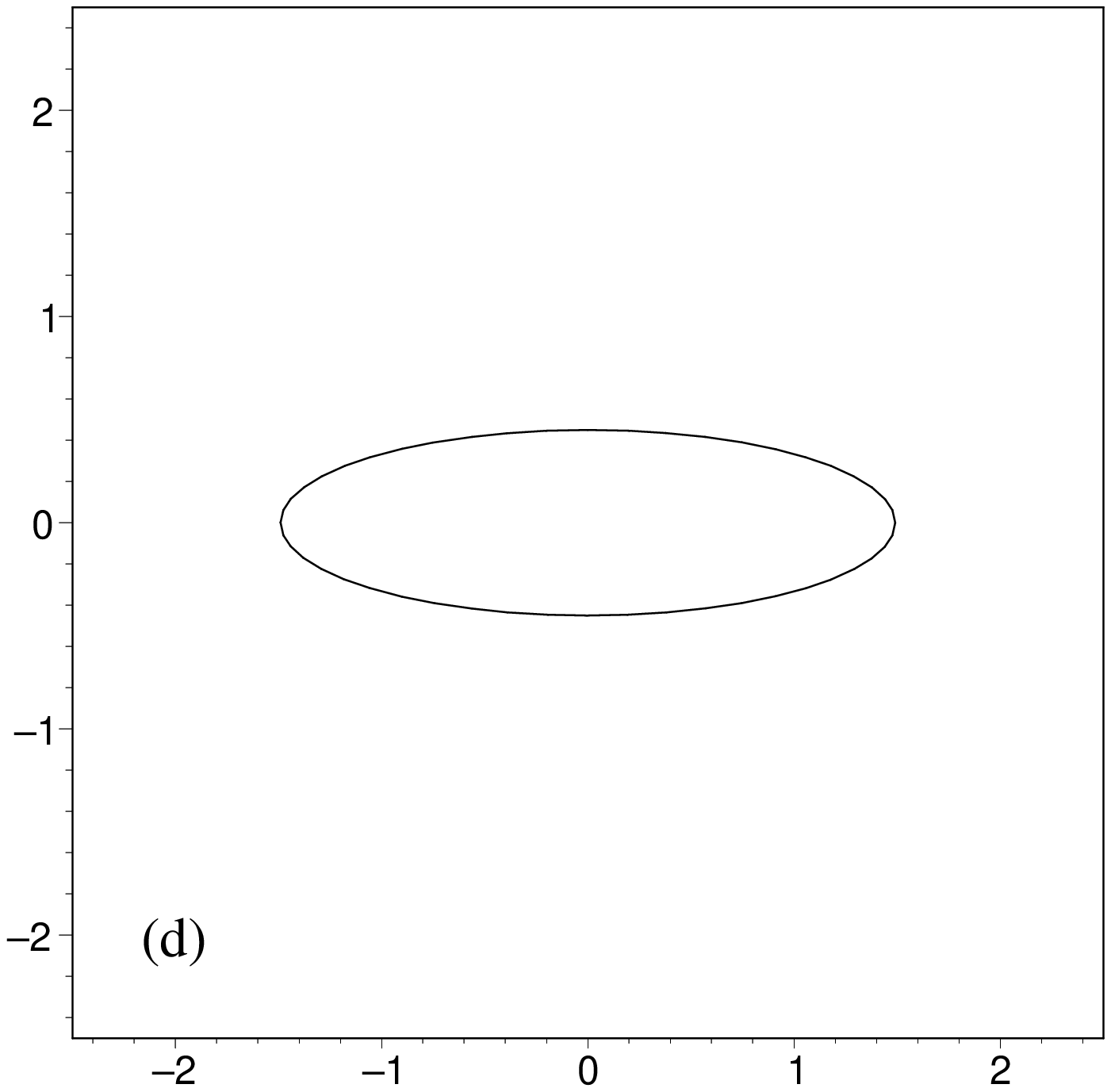}&\\[.2cm]
\includegraphics[scale=0.4]{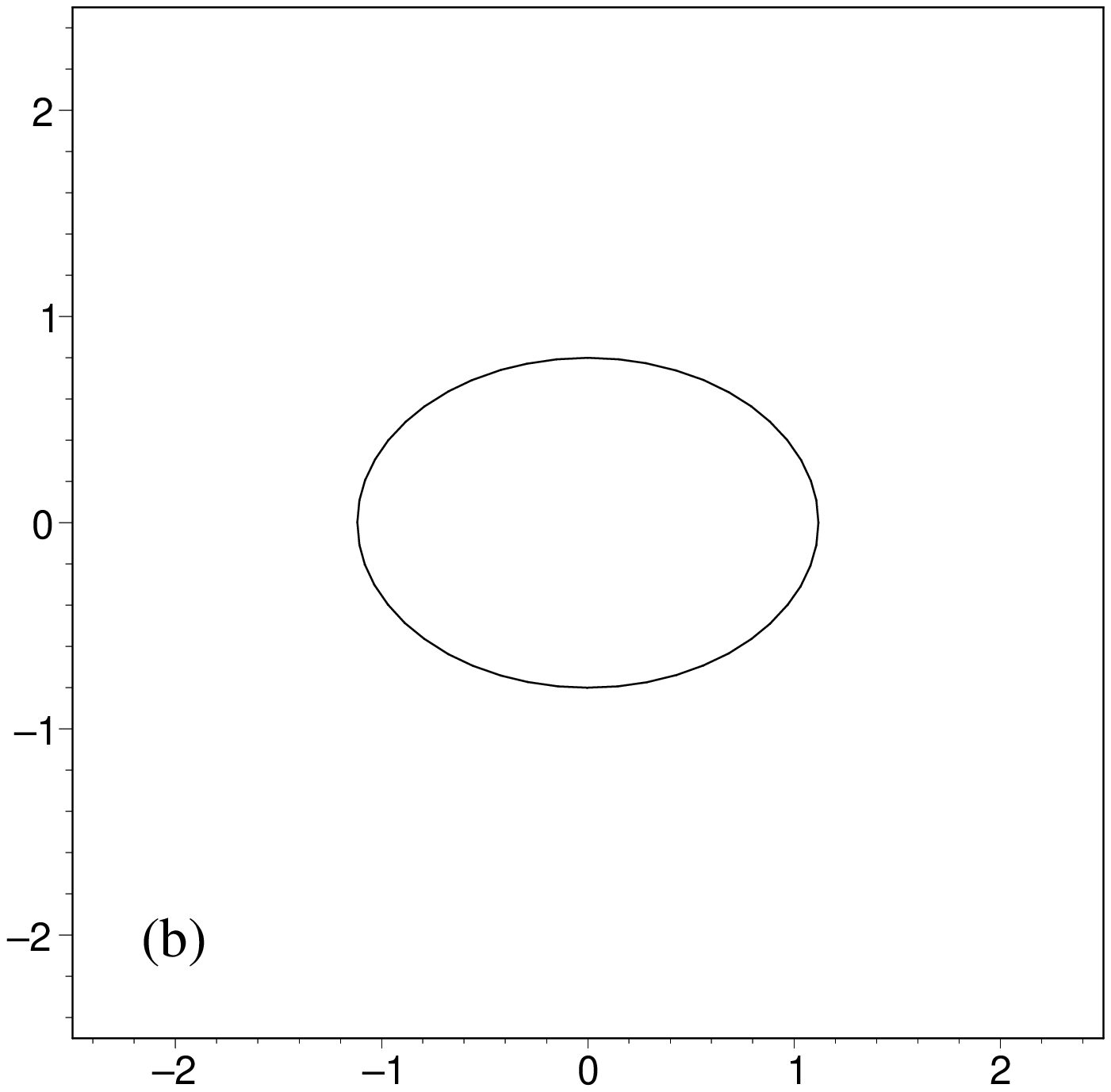}&\quad
\includegraphics[scale=0.4]{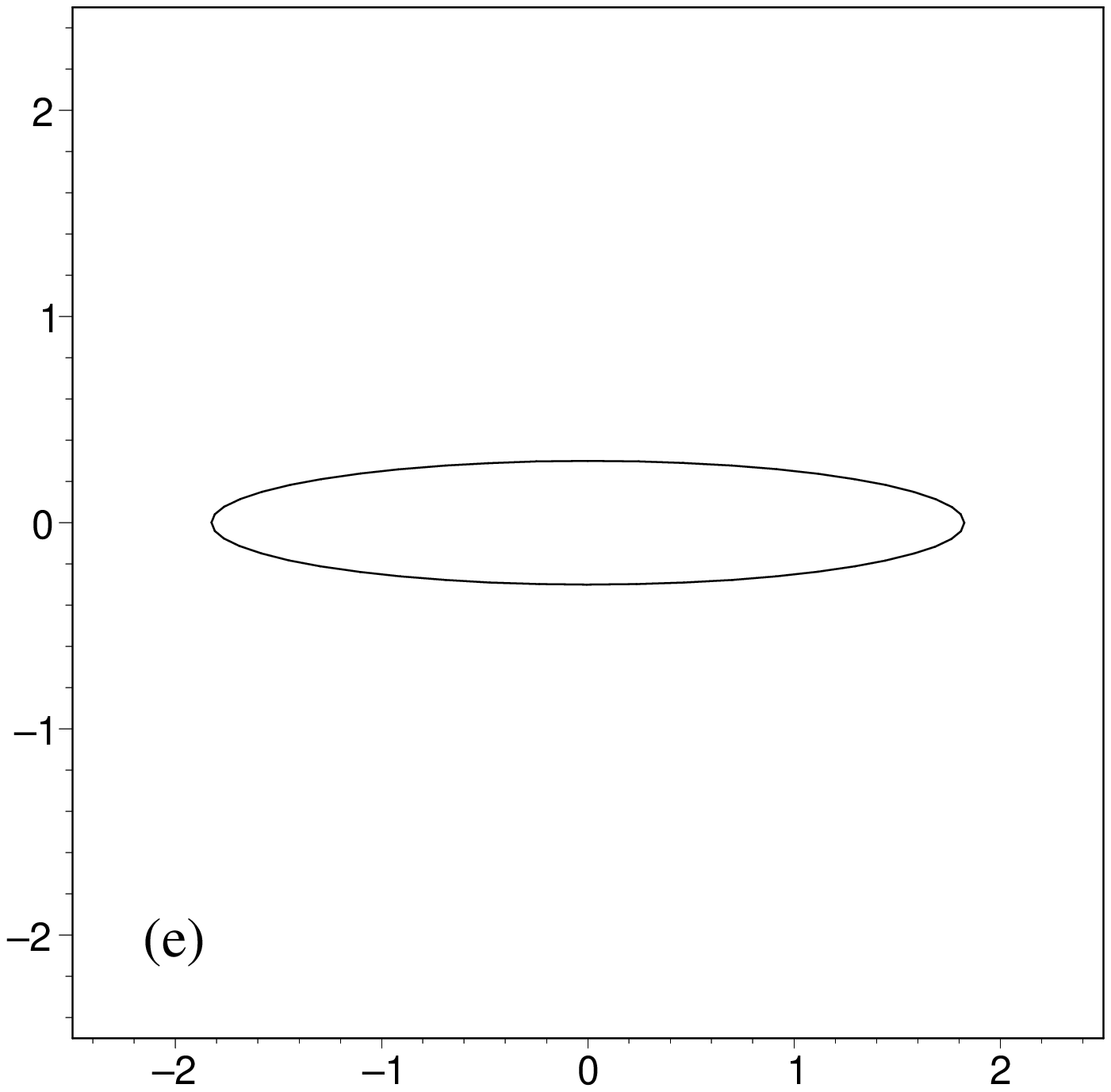}\\[.2cm]
\includegraphics[scale=0.4]{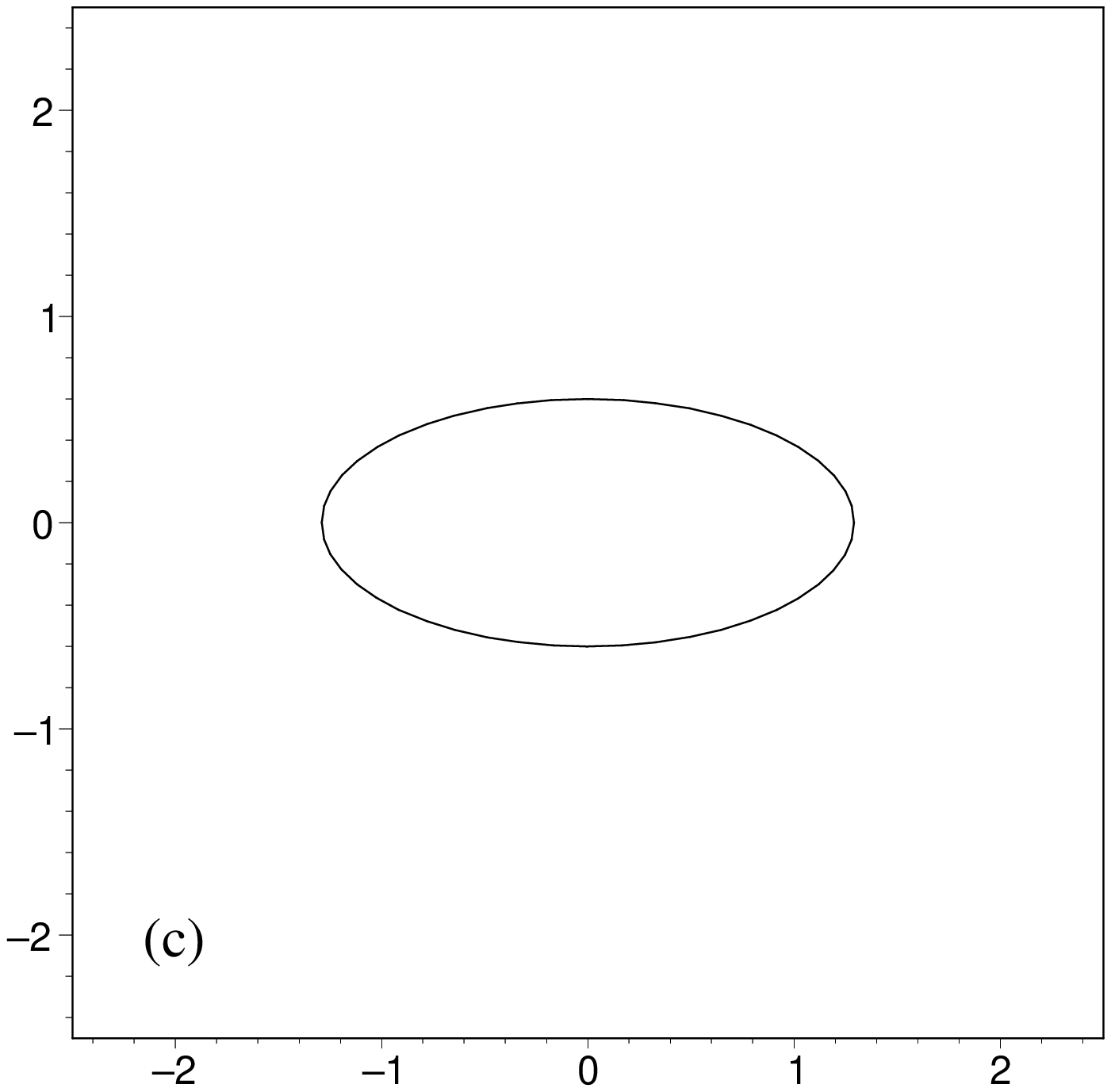}&\quad
\includegraphics[scale=0.4]{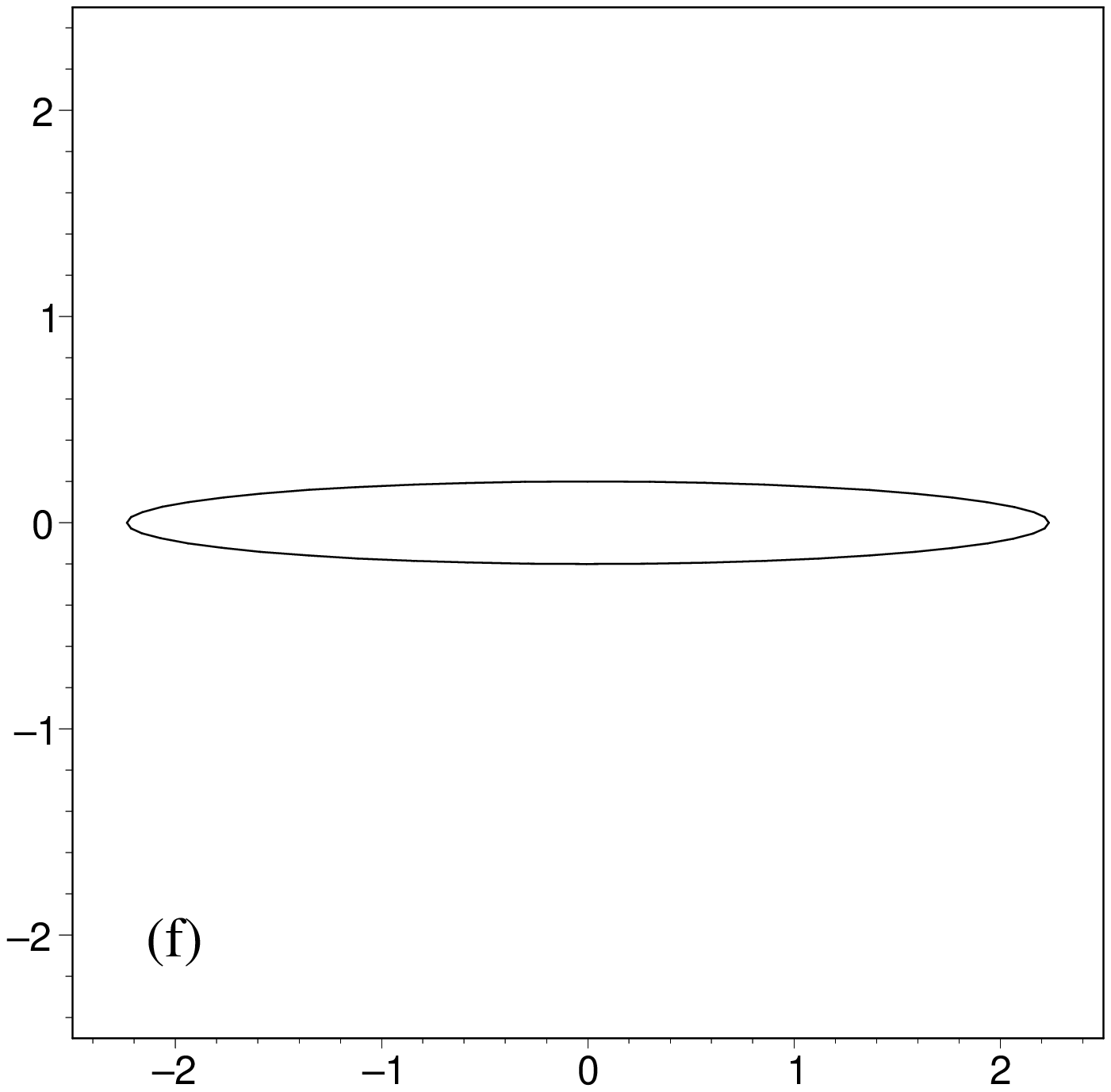}
\end{array}
$\\
\end{center}
\caption{Kerr spacetime, Painlev\'e-Gullstrand observers. An initially circular bunch of particles on the $Y'{}^1$-$Y'{}^2$ plane at $r_0/\mathcal{M}=10$ and $\theta=\pi/2$ (see Fig.~(a)) is squeezed along the radial direction. 
The curves depicted in Figs.~(b) to (f) correspond to decreasing values of the radial coordinate $r/\mathcal{M}=[10,8,6,4.5,3,2]$ and increasing values of the proper time $\tau_{\mathcal{N}}/\mathcal{M}=[0,4.23,7.96,10.38,12.41,13.51]$ (for the additional choice $a/\mathcal{M}=.5$), respectively. 
}
\label{fig:7}
\end{figure}

The situation greatly simplifies on the equatorial plane ($\theta=\pi/2$), where the deviation matrix becomes diagonal
\beq
{\mathcal K}_{({\mathcal N},E')}=\frac{\mathcal{M}}{r^3}{\rm diag}\left[-2-\frac{3a^2}{r^2},1+\frac{3a^2}{r^2},1\right]\ ,
\eeq
and the only nonvanishing components of the expansion tensor are given by
\beq\fl\quad
[\theta'(\mathcal{N})_{11},\theta'(\mathcal{N})_{22},\theta'(\mathcal{N})_{33}]=\sqrt{\frac{2\mathcal{M}}{r^3}}\left[\frac12\frac{\sqrt{r^2+a^2}}{r},-\frac{\sqrt{r^2+a^2}}{r},-\frac{r}{\sqrt{r^2+a^2}}\right]\ .
\eeq
The system (\ref{eq1ordPGfw}) then writes as 
\beq
\label{eq1ordPGfweq}
\frac{\rmd Y'{}^1}{\rmd r}=-\frac{Y'{}^1}{2r}\ , \quad 
\frac{\rmd Y'{}^2}{\rmd r}=\frac{Y'{}^2}{r}\ , \quad 
\frac{\rmd Y'{}^3}{\rmd r}=Y'{}^3\frac{r}{r^2+a^2}\ ,  
\eeq
and can be easily integrated:
\beq
\label{sol4383}
Y'{}^1=Y'{}^1_0\sqrt{\frac{r_0}{r}}\ , \quad 
Y'{}^2=Y'{}^2_0\frac{r}{r_0}\ , \quad 
Y'{}^3=Y'{}^3_0\sqrt{\frac{r^2+a^2}{r_0^2+a^2}}\ . 
\eeq
Note that the solution (\ref{sol4383}) for $Y'{}^3$ is valid everywhere, not only on the equatorial plane, since Eq.~(\ref{eq1ordPGfweq})$_3$ is actually general.
The behaviours of the components $Y'{}^a$ as functions of $r$ are practically the same as those shown in Fig.~\ref{fig:6}.
Figure \ref{fig:7} shows instead the squeezing of an initially circular bunch of particles on the $Y'{}^1$-$Y'{}^2$ plane for decreasing values of the radial coordinate. 
No significative differences appear for different values of the polar angle $\theta$; the squeezing on the $Y'{}^1$-$Y'{}^3$ plane also exhibits the same features although with a slight but not negligible difference from the $Y'{}^2$ component at small values of $r$ as can be seen from Fig.~\ref{fig:6}.


We notice that the Painlev\'e-Gullstrand  family of geodesics may be considered suitable to describe free fall type of accretion by a set of rotationally dragged particles. 

\subsection{Circular orbits in stationary axisymmetric spacetimes}

Consider a family $U$ of circular orbits at different radii around a certain reference value $r_0$, all rotating with the same angular velocity, which we also assume to be constant.

The constant angular velocity ensures that $U$ is a Born-rigid congruence: $\theta(U)=0$, that is $K(U)=\omega(U)$ (vorticity tensor) and $K(U)\rightcontract Y=-\omega(U)\times Y$ (vorticity vector, with an abuse of notation).
Let us use a Frenet-Serret triad $\{E_a\}$ along each $U$ of the family, which rotates with respect to a Fermi-Walker transported frame along $U$ with angular velocity
\beq
\label{omegaFS}
\omega_{\rm (FS)}=\tau_1 E_3 + \tau_2 E_1\ , \qquad ||\omega_{\rm (FS)}||=[\tau_1^2 + \tau_2^2]^{1/2}\ ,
\eeq
according to 
\beq
\frac{DE_a}{d\tau_U}=\omega_{\rm (FS)}\times E_a+\kappa E_0\,\delta^1_a\ .
\eeq
The curvature $\kappa$ as well as the first and second torsions $\tau_1$ and $\tau_2$ are constant along the orbit in this case \cite{iyer-vish}.
It results that $\omega_{ ({\rm fw},U,E)}=\omega_{\rm (FS)}=\omega(U)$, that is the Frenet-Serret angular velocity coincides with the vorticity of the congruence.
Moreover, the alignment of the orbit's four velocity with a Killing vector implies
\beq
\dot \omega_{ ({\rm fw},U,E)}=0\ ,
\eeq
so that Eq.~(\ref{eq:secder2}) simplifies to
\begin{eqnarray}
\label{eq:secder3}
\nabla_{UU} Y-\nabla_Ya(U)
&=&  -Y^b\nabla(U)_b a(U)+\omega(U)\times [\omega(U)\times Y]+\nonumber \\
&& +\ddot Y^a\, E_a-[Y\cdot a(U)]a(U)\ .
\end{eqnarray}
It can be shown that the electric part of the Weyl tensor for vacuum stationary spacetimes is given by \cite{mfg}
\beq
\fl\quad
{\mathcal E}(U)=-\omega(U) \otimes \omega(U) +  a(U)\otimes a(U) + \nabla (U)a(U) +\omega(U)^2\, P(U)\ ,
\eeq
whence
\beq
\fl\quad
-{\mathcal E}(U)\rightcontract Y= \omega(U)\times [\omega(U)\times Y]-[a(U) \cdot Y]a(U)-Y^b\nabla (U)_ba(U)\ .
\eeq
Equating both sides of the previous equation with the corresponding ones of Eq.~(\ref{eq:secder3}) leads to
\beq
\label{der2Y}
\ddot Y^a\equiv 0\ .
\eeq

The same result can be obtained by direct evaluation of the Lie transport equation (\ref{eq:1})$_2$ for the connecting vector $Y$, which in the FS frame reads 
\beq
\label{FSlieeqs}
\dot Y^0=0\ , \qquad \dot Y^a+[(\omega_{\rm (FS)}-\omega(U))\times Y]^a-[\theta(U)\rightcontract Y]^a=0\ . 
\eeq
Since $\theta(U)=0$ and $\omega_{\rm (FS)}=\omega(U)$ in this case, the previous equations simplify to 
\beq
\dot Y^0=0\ , \qquad \dot Y^a=0\ , 
\eeq
trivially implying Eq.~(\ref{der2Y}).

This case is the most interesting since it naturally describes the orbital motion of a system which we may require to remain rigid as for example an artificial satellite around the Earth or a \lq\lq planet" orbiting a star. The physical constraints which assure rigidity and therefore stability of the orbiting system are deduced directly from Eq.~(\ref{FSlieeqs}).

Finally, one can set up a Fermi-Walker transported frame adapted to $U$ also in this case, following exactly the same procedure as in Section \ref{thread}. It will result that the component of the deviation vector along the Fermi-Walker angular velocity remains constant along the path, while the other spatial components oscillate with the same frequency $||\omega_{({\rm fw},U,E)}||\equiv||\omega_{\rm (FS)}||$.

\subsection{Schwarzschild spacetime limit}

In the case of vanishing rotation parameter the family of ZAMOs coincides with that of static observers, $n\equiv m$.
The expansion tensor $\theta(n)$ as well as the Fermi-Walker angular velocity $\omega_{({\rm fw},n,E)}{}$ are identically zero, whence the matrix 
$T_{({\rm fw},n,E)}\equiv 0$ too.
Furthermore we have that
\beq
{\mathcal E}(n)=S(n)=\frac{\mathcal{M}}{r^3}{\rm diag}[-2,1,1]\ ,
\eeq
implying that also the deviation matrix ${\mathcal K}_{(n,E)}$ is identically zero.
Hence the spatial components of the deviation vector remain all constant along the path, since both $\ddot Y^a=0$ and $\dot Y^a=0$ (note that they are only functions of $r$).

In the case of Painlev\'e-Gullstrand observers, instead, the limit of vanishing rotation parameter leads to the radial geodesic motion. The matrix ${\mathcal K}_{({\mathcal N},E)}$ results to be diagonal and completely determined in terms of the surviving components of the electric part of the Weyl tensor
\beq
{\mathcal K}_{({\mathcal N},E)}={\mathcal E}(\mathcal {N})=\frac{\mathcal{M}}{r^3}{\rm diag}[-2,1,1]\ .
\eeq
As a result, the system of deviation equations (\ref{eq:secder3bis}) decouples as follows:
\beq\fl\quad
\ddot Y^1+{\mathcal K}_{({\mathcal N},E)}{}_{11}Y^1=0\ , \quad
\ddot Y^2+{\mathcal K}_{({\mathcal N},E)}{}_{22}Y^2=0\ , \quad
\ddot Y^3+{\mathcal K}_{({\mathcal N},E)}{}_{33}Y^3=0\ , 
\eeq
where the coefficients ${\mathcal K}_{({\mathcal N},E)}{}_{11},{\mathcal K}_{({\mathcal N},E)}{}_{22},{\mathcal K}_{({\mathcal N},E)}{}_{33}$ are functions of the proper time $\tau_\mathcal{N}$.
We notice that the frame (\ref{framePG}) is already Fermi-Walker transported also in this case, since $\omega_{({\rm fw},{\mathcal N},E)}\equiv 0$; moreover, the Fermi-Walker triad (\ref{fwPG}) reduces to the spatial triad of (\ref{framePG}) in the limit $a=0$.
The corresponding first order system of Lie transport equations (\ref{eq1ordPG}) thus simplifies to 
\beq
\label{schwPG1ord}
\dot Y^1=\theta(\mathcal{N})_{11}Y^1\ , \quad 
\dot Y^2=\theta(\mathcal{N})_{22}Y^2\ , \quad 
\dot Y^3=\theta(\mathcal{N})_{33}Y^3\ ,
\eeq
where
\beq
[\theta(\mathcal{N})_{11},\theta(\mathcal{N})_{22},\theta(\mathcal{N})_{33}]=\sqrt{\frac{2\mathcal{M}}{r^3}}\left[\frac12,-1,-1\right]\ ,
\eeq
from Eq.~(\ref{thetaPG}), and can be easily integrated. 
After introducing the explicit dependence on the radial coordinate instead of the proper time parametrization Eqs.~(\ref{schwPG1ord}) then become
\beq
\frac{\rmd Y^1}{\rmd r}=-\frac{Y^1}{2r}\ , \quad 
\frac{\rmd Y^2}{\rmd r}=\frac{Y^2}{r}\ , \quad 
\frac{\rmd Y^3}{\rmd r}=\frac{Y^3}{r}\ ,  
\eeq
whose solution is straightforward:
\beq
Y^1=Y^1_0\sqrt{\frac{r_0}{r}}\ , \quad 
Y^2=Y^2_0\frac{r}{r_0}\ , \quad 
Y^3=Y^3_0\frac{r}{r_0}\ . 
\eeq
In contrast with the Kerr case the deviations along the angular directions $Y^2$ and $Y^3$ have thus exactly the same behaviour as functions of the radial coordinate.
As a result, the squeezing of an initially circular bunch of particles on the $Y^1$-$Y^2$ plane is the same as the one on the $Y^1$-$Y^3$ plane; its shape changes for decreasing values of the radial coordinate in a way very similar to the corresponding one for Kerr, as shown in Fig.~\ref{fig:7}. 

\section{Conclusions}

We have defined relative accelerations and strains among a set of comoving particles (nongeodesic congruence of timelike world lines) with respect to a \lq\lq fiducial observer" world line (i.e. a single reference line of the congruence) in terms of the geometric properties of both the congruence and the fiducial observers.
We have provided an operational definition of strains, based on spacetime splitting techniques, evidentiating the role of the observer as well as the choice of the spatial triad associated with it, otherwise arbitrary and arbitrarily dragged along the observer world line. Strains and deviation vectors may have strongly different behaviours according to the preferred reference frame set up for their measurements, a point that is somehow missing in the literature. 
A first attempt due to Szekeres as well as further developments (more recent) due to Mashhoon and coworkers (concerning the theory of the relativistic gravity gradiometer) and de Felice and coworkers (concerning the definition of strains) have been encompassed. 
To better specify the results of our analysis we have studied certain special congruences in the Kerr spacetime (static observers, ZAMOs, Painlev\'e-Gullstrand, circular orbits). 
Particular attention has been devoted to the Fermi-Walker frame, since it represents a reference system operationally defined in terms of three gyroscopes dragged along the observer world line. In the Kerr case, when using Fermi-Walker axes, static observers and ZAMOs experience certain relative strains which result in harmonic oscillations for the deviation vectors; differently, the Painlev\'e-Gullstrand observers (free-falling and locally nonrotating, spiraling towards the singularity) feel decreasing deviation along the angular directions (approaching the singularity), while increasing deviation in the radial direction. 
This analysis can be repeated in other interesting gravitational situations and it should be very useful when thinking about general relativistic experiments on space stations in black hole gravitational backgrounds.

\appendix
\section{Kerr spacetime: relevant quantities}

We list below the relevant quantities entering the relative acceleration equation (\ref{eq:secder3bis}) and corresponding to all families of observers congruences considered in Section \ref{kerrcase} for the case of a Kerr spacetime.

\subsection{Static observers}
\label{appKthread}

The components of the electric part of the Weyl tensor as well as the relevant kinematical quantities entering Eq.~(\ref{eq:secder3bis}) and corresponding to a congruence of static observers are given by

\begin{eqnarray}
\label{variousqua_thd}
\fl\quad
{\mathcal E}(m)_{11}&=&\frac{\mathcal{M}r}{\Sigma^3}(4r^2-3\Sigma)  \frac{2\Delta+a^2\sin^2\theta}{2\mathcal{M}r-\Sigma}\ , \nonumber\\
\fl\quad
{\mathcal E}(m)_{12}&=&-\frac{3\mathcal{M}a^2\sqrt{\Delta}}{\Sigma^3}(4r^2-\Sigma)\frac{\cos\theta\sin\theta}{2\mathcal{M}r-\Sigma}\ , \nonumber\\
\fl\quad
{\mathcal E}(m)_{22}&=&-\frac{\mathcal{M}r}{\Sigma^3}(4r^2-3\Sigma)\frac{\Delta+2a^2\sin^2\theta}{2\mathcal{M}r-\Sigma}\ , \nonumber\\
\fl\quad
{\mathcal E}(m)_{33}&=&\frac{\mathcal{M}r}{\Sigma^3}(4r^2-3\Sigma)\ , \nonumber\\
\fl\quad
\omega(m)_{13}&=&\frac{a\mathcal{M}}{\Sigma^{3/2}}(2r^2-\Sigma)\frac{\sin\theta}{2\mathcal{M}r-\Sigma}
=-\omega_{({\rm fw},m,E)}{}_{2}\ , \nonumber\\
\fl\quad
\omega(m)_{23}&=&\frac{2a\mathcal{M}r\sqrt{\Delta}}{\Sigma^{3/2}}\frac{\cos\theta}{2\mathcal{M}r-\Sigma}
=\omega_{({\rm fw},m,E)}{}_{1}\ , \nonumber\\
\fl\quad
T_{({\rm fw},m,E)}{}_{11}&=&-\omega_{({\rm fw},m,E)}{}_{2}^2\ , \quad T_{({\rm fw},m,E)}{}_{12}=\omega_{({\rm fw},m,E)}{}_{1}\omega_{({\rm fw},m,E)}{}_{2}\ , \nonumber\\
\fl\quad
T_{({\rm fw},m,E)}{}_{22}&=&-\omega_{({\rm fw},m,E)}{}_{1}^2\ , \quad T_{({\rm fw},m,E)}{}_{33}=-||\omega_{({\rm fw},m,E)}||^2\ .
\end{eqnarray}
We recall that all the components of the tensor fields in Eq.~(\ref{variousqua_thd}) refer to the (orthonormal) adapted frame (\ref{thdframe}).

By using Eq.~(\ref{variousqua_thd}) it results $S(m)={\mathcal E}(m)+T_{({\rm fw},m,E)}{}$, implying that the matrix ${\mathcal K}_{(m,E)}$ is identically zero.

\subsection{ZAMOs}
\label{appKZAMO}

The components of the electric part of the Weyl tensor as well as the relevant kinematical quantities entering Eq.~(\ref{eq:secder3bis}) and corresponding to a congruence of ZAMOs are given by
\begin{eqnarray}\fl\quad
\label{variousqua_zamo}
{\mathcal E}(n)_{11}&=&-\frac{\mathcal{M}r}{\Sigma^3}(4r^2-3\Sigma)\frac{2(r^2+a^2)^2+a^2\Delta\sin^2\theta}{(r^2+a^2)^2-a^2\Delta\sin^2\theta}\ , \nonumber\\
\fl\quad
{\mathcal E}(n)_{12}&=&\frac{3a^2\mathcal{M}\sqrt{\Delta}}{\Sigma^3}(4r^2-\Sigma)\frac{(r^2+a^2)\cos\theta\sin\theta}{(r^2+a^2)^2-a^2\Delta\sin^2\theta}\ , \nonumber\\
\fl\quad
{\mathcal E}(n)_{22}&=&\frac{\mathcal{M}r}{\Sigma^3}(4r^2-3\Sigma)\frac{(r^2+a^2)^2+2a^2\Delta\sin^2\theta}{(r^2+a^2)^2-a^2\Delta\sin^2\theta}\ , \nonumber\\
\fl\quad
{\mathcal E}(n)_{33}&=&\frac{\mathcal{M}r}{\Sigma^3}(4r^2-3\Sigma)\ , \nonumber\\
\fl\quad
\theta(n)_{13}&=&-\frac{a\mathcal{M}}{\Sigma^{3/2}}\frac{(r^2-a^2)\Sigma +2r^2 (r^2+a^2)}{(r^2+a^2)^2-a^2\Delta\sin^2\theta }\sin\theta
=\omega_{({\rm fw},n,E)}{}_{2}\ , \nonumber\\
\fl\quad
\theta(n)_{23}&=&\frac{2a^3\mathcal{M}r\sqrt{\Delta}}{\Sigma^{3/2}}\frac{\cos\theta\sin^2\theta}{(r^2+a^2)^2-a^2\Delta\sin^2\theta}
=-\omega_{({\rm fw},n,E)}{}_{1}\ , \nonumber\\
\fl\quad
T_{({\rm fw},n,E)}{}_{11}&=&3\omega_{({\rm fw},n,E)}{}_{2}^2\ , \quad T_{({\rm fw},n,E)}{}_{12}=-3\omega_{({\rm fw},n,E)}{}_{1}\omega_{({\rm fw},n,E)}{}_{2}\ , \nonumber\\
\fl\quad
T_{({\rm fw},n,E)}{}_{22}&=& 3\omega_{({\rm fw},n,E)}{}_{1}^2\ , \quad T_{({\rm fw},n,E)}{}_{33}=-||\omega_{({\rm fw},n,E)}||^2\ .
\end{eqnarray}
We recall that all the components of the tensor fields in Eq.~(\ref{variousqua_zamo}) refer to the (orthonormal) adapted frame (\ref{zamoframe}).

By using Eq.~(\ref{variousqua_zamo}) it results $S(n)={\mathcal E}(n)+T_{({\rm fw},n,E)}$, implying that the matrix ${\mathcal K}_{(n,E)}$ is identically zero.

\subsection{Painlev\'e-Gullstrand observers}
\label{appKPG}

The kinematical properties of the $\mathcal{N}$ observers are summarized by the expansion
$\theta(\mathcal{N})$. In fact the exterior derivative of (\ref{PG4vel}) is zero: $\rmd \mathcal{N}=0$, implying that $a(\mathcal{N})=0$ and $\omega(\mathcal{N})=0$ \cite{mfg}.
In this case $S(\mathcal{N})=0$, so that ${\mathcal K}_{({\mathcal N},E)}=T_{({\rm fw},\mathcal{N},E)}+{\mathcal E}(\mathcal{N})$ and in addition 
$T_{({\rm fw},\mathcal{N},E)}$ has the form
\begin{eqnarray}
\fl\quad
T_{({\rm fw},\mathcal{N},E)}{}^a{}_b&=&\delta^a_b  \omega_{({\rm fw},\mathcal{N},E)}^2- \omega_{({\rm fw},\mathcal{N},E)}^a 
 \omega_{({\rm fw},\mathcal{N},E)}{}_b \nonumber \\
\fl\quad
&& -\epsilon^a{}_{bf}\dot  \omega_{({\rm fw},{\mathcal N},E)}^f+2\epsilon^a{}_{fc}  \omega_{({\rm fw},\mathcal{N},E)}^f \theta(\mathcal{N})^c{}_b\ . 
\end{eqnarray}
The nonvanishing frame  components of the expansion tensor are given by
\begin{eqnarray}\fl
\label{thetaPG}
\theta(\mathcal {N})_{11}&=&-\gamma(\mathcal{N},n)\nu(\mathcal{N},n)\frac{\sqrt{\Delta}}{2r\Sigma^{1/2}(r^2+a^2)}\frac{\Sigma(r^4-a^4)+2r^2a^2\Delta\sin^2\theta}{(r^2+a^2)^2-a^2\Delta\sin^2\theta}\ , \nonumber\\
\fl
\theta(\mathcal {N})_{12}&=&\frac{a^2}{\Sigma^{3/2}}\nu(\mathcal{N},n)\sin\theta\cos\theta\ , \nonumber\\
\fl
\theta(\mathcal {N})_{13}&=&-\frac{a\mathcal{M}}{\Sigma^{3/2}}\frac{(r^2-a^2)\Sigma+2r^2(r^2+a^2)}{(r^2+a^2)^2-a^2\Delta\sin^2\theta}\sin\theta\ \ , \nonumber\\
\fl
\theta(\mathcal {N})_{22}&=&\gamma(\mathcal{N},n)\nu(\mathcal{N},n)\frac{r\sqrt{\Delta}}{\Sigma^{3/2}}\ , \nonumber\\
\fl
\theta(\mathcal {N})_{23}&=&\gamma(\mathcal{N},n)\frac{2a^3\mathcal{M}r\sqrt{\Delta}}{\Sigma^{3/2}}\frac{\cos\theta\sin^2\theta}{(r^2+a^2)^2-a^2\Delta\sin^2\theta}\ , \nonumber\\
\fl
\theta(\mathcal {N})_{33}&=&\gamma(\mathcal{N},n)\nu(\mathcal{N},n)\frac{\sqrt{\Delta}}{\Sigma^{3/2}}\frac{(r-\mathcal{M})\Sigma^2+\mathcal{M}(3r^2+a^2)\Sigma-2\mathcal{M}r^2(r^2+a^2)}{(r^2+a^2)^2-a^2\Delta\sin^2\theta}\ ,
\end{eqnarray}
and for the components of the angular velocity vector $ \omega_{({\rm fw},\mathcal{N},E)}$ we find
\beq\fl\quad
\label{omfwPG}
 \omega_{({\rm fw},\mathcal{N},E)}{}_1=-\theta(\mathcal {N})_{23}\ , \quad 
 \omega_{({\rm fw},\mathcal{N},E)}{}_2=\theta(\mathcal {N})_{13}\ , \quad
 \omega_{({\rm fw},\mathcal{N},E)}{}_3=\theta(\mathcal {N})_{12}\ .
\eeq
We notice that the electric part of the Riemann tensor associated with the Painlev\'e-Gullstrand observers can be obtained using the general transformation laws mapping the fiels measured by a family of observers into the corresponding ones measured by another family \cite{mfg}.
In the special case of vacuum spacetimes, using the decomposition (\ref{calN}) it is easy to show that
\begin{eqnarray}
\label{trasfriem}
\fl\quad
{\mathcal E}({\mathcal N})_{\alpha\beta} &=& \gamma({\mathcal N},n)^2 [P(n,{\mathcal N})^{-1}]_\alpha{}^\mu [P(n,{\mathcal N})^{-1}]_\beta{}^\nu \bigl[
{\mathcal E}(n)_{\mu\nu}-2 {\mathcal H}(n)_{(\mu|\rho} V^\rho{}_{\nu )}+\nonumber \\
\fl\quad
&&+V_\mu{}^\rho {\mathcal E}(n)_{\rho\sigma} V^\sigma{}_\nu \bigr]\ ,
\end{eqnarray}
with the analogous expression for ${\mathcal H}({\mathcal N})$ (simply obtained with the replacements: ${\mathcal E} \to {\mathcal H}$ and 
${\mathcal  H}\to -{\mathcal E}$). 
Here $[P(n,{\mathcal N})^{-1}]_\alpha{}^\beta=P(n)_\alpha{}^\beta+n_\alpha \nu({\mathcal N},n)^\beta$, with $\nu({\mathcal N},n)^\beta=\nu({\mathcal N},n)\,\delta^\beta_{\hat r}$, projects from the ZAMO local rest space to that of the Painlev\'e-Gullstrand observers, and we have used the notation
\beq
V_{\alpha\beta}\equiv V({\mathcal N},n)=\eta(n)_{\alpha\beta\gamma}\nu({\mathcal N},n)^\gamma\ , \quad
\eta(n)_{\alpha\beta\gamma}=n^\mu \eta_{\mu\alpha\beta\gamma}\ ,
\eeq
together with the standard definitions
\begin{eqnarray}
\label{elmamis}
{\mathcal E}(n)_{\beta\delta} &=& R_{\alpha\beta\gamma\delta}n^\alpha 
n^\gamma \ , \quad 
{\mathcal H}(n)_{\beta\delta} =-R^*_{\alpha\beta\gamma\delta} n^\alpha 
n^\gamma \ , 
\end{eqnarray}
for the electric and magnetic part of the Riemann tensor with respect to the ZAMO family of observers, given by Eq.~(\ref{variousqua_zamo}) and 
\begin{eqnarray}\fl\quad
&&{\mathcal H}(n)_{11}=k{\mathcal E}(n)_{11}\ , \quad {\mathcal H}(n)_{12}=-k^{-1}{\mathcal E}(n)_{12}\ , \quad {\mathcal H}(n)_{22}=k{\mathcal E}(n)_{22}\ , \nonumber\\
\fl\quad
&&{\mathcal H}(n)_{33}=k{\mathcal E}(n)_{33}\ , \quad k=\frac{a}{r}\frac{4r^2-\Sigma}{4r^2-3\Sigma}\cos\theta\ ,
\end{eqnarray}
respectively. 
The long formulas for the components of ${\mathcal E}({\mathcal N})$ are listed below:
\begin{eqnarray}\fl\quad
{\mathcal E}(\mathcal {N})_{11}&=&-\frac{\mathcal{M}r}{\Sigma^3}(4r^2-3\Sigma)\frac{2(r^2+a^2)^2+a^2\Delta\sin^2\theta}{(r^2+a^2)^2-a^2\Delta\sin^2\theta}\ , \nonumber\\
\fl\quad
{\mathcal E}(\mathcal {N})_{12}&=&\gamma(\mathcal{N},n)\frac{3\mathcal{M}a^2\sqrt{\Delta}}{\Sigma^3}(4r^2-\Sigma)\frac{(r^2+a^2)\cos\theta\sin\theta}{(r^2+a^2)^2-a^2\Delta\sin^2\theta}\ , \nonumber\\
\fl\quad
{\mathcal E}(\mathcal {N})_{13}&=&-\gamma(\mathcal{N},n)\nu(\mathcal{N},n)\frac{3a\mathcal{M}r\sqrt{\Delta}}{\Sigma^3}\frac{(4r^2-3\Sigma)(r^2+a^2)}{(r^2+a^2)^2-a^2\Delta\sin^2\theta}\sin\theta\ , \nonumber\\
\fl\quad
{\mathcal E}(\mathcal {N})_{22}&=&\frac{\mathcal{M}r}{\Sigma^4}(4r^2-3\Sigma)[3(r^2+a^2)-2\Sigma]\ , \nonumber\\
\fl\quad
{\mathcal E}(\mathcal {N})_{23}&=&\nu(\mathcal{N},n)\frac{3a^3\mathcal{M}}{\Sigma^4}(4r^2-\Sigma)\cos\theta\sin^2\theta\ , \nonumber\\
\fl\quad
{\mathcal E}(\mathcal {N})_{33}&=&-\frac{\mathcal{M}r}{\Sigma^4}(4r^2-3\Sigma)\frac{-\Sigma^2\Delta+2\mathcal{M}r[3(r^2+a^2)-4\Sigma]}{(r^2+a^2)^2-a^2\Delta\sin^2\theta}\ .
\end{eqnarray}
Using spacetime splitting techniques \cite{mfg}, it is also possible to show that
\begin{eqnarray}
{\mathcal E}(\mathcal {N})_{\alpha\beta}&=&[-\nabla_{\rm (lie) }(\mathcal{N}) \theta(\mathcal{N})+\theta(\mathcal{N})^2]^{\rm (TF)}{}_{\alpha\beta}\ ,
\end{eqnarray}
where $\nabla_{\rm (lie) }(\mathcal{N})=P(\mathcal{N})\pounds_{\mathcal{N}}$ is the the projected Lie temporal derivative and ${\rm TF}$ denotes the trace-free part of a tensor.
The nonvanishing components of ${\mathcal K}_{(\mathcal N,E)}$ are then given by
\begin{eqnarray}\fl\quad
\label{KabPG}
{\mathcal K}_{({\mathcal N},E)}{}_{11}&=&-\dot\theta(\mathcal {N})_{11}-\theta(\mathcal {N})_{11}^2\ , \nonumber\\
\fl\quad 
{\mathcal K}_{({\mathcal N},E)}{}_{12}&=&-2[\dot\omega_{({\rm fw},\mathcal{N},E)}{}_3+\omega_{({\rm fw},\mathcal{N},E)}{}_3(\theta(\mathcal {N})_{11}+\theta(\mathcal {N})_{22})]\ , \nonumber\\
\fl\quad 
{\mathcal K}_{({\mathcal N},E)}{}_{22}&=&-\dot\theta(\mathcal {N})_{22}-\theta(\mathcal {N})_{22}^2\ , \nonumber\\
\fl\quad 
{\mathcal K}_{({\mathcal N},E)}{}_{31}&=&-2[\dot\omega_{({\rm fw},\mathcal{N},E)}{}_2+\omega_{({\rm fw},\mathcal{N},E)}{}_2(\theta(\mathcal {N})_{11}+\theta(\mathcal {N})_{33})]\ , \nonumber\\
\fl\quad 
{\mathcal K}_{({\mathcal N},E)}{}_{32}&=&2[\dot\omega_{({\rm fw},\mathcal{N},E)}{}_1+\omega_{({\rm fw},\mathcal{N},E)}{}_1(\theta(\mathcal {N})_{22}+\theta(\mathcal {N})_{33})\nonumber \\
\fl\quad
&&-2\omega_{({\rm fw},\mathcal{N},E)}{}_2\omega_{({\rm fw},\mathcal{N},E)}{}_3]\ , \nonumber\\
\fl\quad 
{\mathcal K}_{(\mathcal N,E)}{}_{33}&=&-\dot\theta(\mathcal {N})_{33}-\theta(\mathcal {N})_{33}^2\ , 
\end{eqnarray}
whose explicit expressions can be easily obtained through Eqs.~(\ref{thetaPG}) and (\ref{omfwPG}).

Finally, we list the nonzero components of the deviation matrix ${\mathcal K}_{({\mathcal N},E')}={\mathcal E}'(\mathcal {N})$ with respect to the Fermi-Walker transported frame (\ref{fwPG})
\begin{eqnarray}
\label{KabPGfw}
\fl\quad
{\mathcal K}_{({\mathcal N},E')}{}_{11}&=&-{\mathcal K}_{({\mathcal N},E')}{}_{22}-{\mathcal K}_{({\mathcal N},E')}{}_{33}\ , \nonumber\\
\fl\quad
{\mathcal K}_{({\mathcal N},E')}{}_{12}&=&\frac{3\mathcal{M}}{a^2r\Sigma^5}\{
-a^4\Sigma^2(\Sigma-4r^2)+\Lambda[r\sqrt{r^2+a^2}(-2\Sigma+2r^2+a^2)\nonumber\\
\fl\quad
&&+a^2(\Sigma-2r^2\sin^2\theta)]\}\sin\theta\cos\theta\ , \nonumber\\
\fl\quad
{\mathcal K}_{({\mathcal N},E')}{}_{22}&=&\frac{\mathcal{M}r}{a^2\Sigma^4}\{
15\Sigma^2(\Sigma-a^2)-4[4r^2(12r^2+5a^2)+3a^4]\Sigma\nonumber\\
\fl\quad
&&+24r^2(8r^2-a^2)(r^2+a^2)\}\nonumber\\
\fl\quad
&&+\frac{3\mathcal{M}\Lambda}{a^2\Sigma^5}\{
2a^2\sin^2\theta[\sqrt{r^2+a^2}\cos^2\theta+r\sin^2\theta]-r(4\Sigma+r^2+a^2)\}\ , \nonumber\\
\fl\quad
{\mathcal K}_{({\mathcal N},E')}{}_{33}&=&\frac{\mathcal{M}r}{\Sigma^3}(4r^2-3\Sigma)\ ,
\end{eqnarray}
where
$$
\Lambda=(5r^2+a^2)\Sigma^2-4r^2(5r^2+3a^2)\Sigma+16r^4(r^2+a^2)\ .
$$

\end{document}